\documentclass[10pt,twocolumn,letterpaper]{article}

\usepackage{iccv}
\usepackage{times}
\usepackage{epsfig}
\usepackage{graphicx}
\usepackage{amsmath}
\usepackage{amssymb}

\usepackage{booktabs}
\usepackage{xcolor}
\usepackage{bbding}

\usepackage{enumitem}
\usepackage{multirow}
\usepackage{bbm}
\usepackage{arydshln}
\usepackage{xspace}
\usepackage{cite}
\usepackage{adjustbox}
\usepackage{booktabs}

\usepackage{times}
\usepackage{epsfig}
\usepackage{graphicx}
\usepackage{amsmath}
\usepackage{amssymb}
\usepackage{makecell}
\usepackage{multirow}
\usepackage{float}
\usepackage{soul}
\usepackage{graphicx}
\usepackage{subcaption}

\usepackage[hang,flushmargin]{footmisc} 

\usepackage[pagebackref,breaklinks,bookmarks=false,colorlinks]{hyperref}

\newcommand{\myparagraph}[1]{{\vspace{.2em} \noindent \bf #1}}

\newcommand*{\affaddr}[1]{#1} 
\newcommand*{\affmark}[1][*]{\textsuperscript{#1}}

\iccvfinalcopy 

\def\iccvPaperID{2933} 

\ificcvfinal\fi

\begin{document}

\title{Semantically Structured Image Compression via \\
 Irregular Group-Based Decoupling}


\author{Ruoyu Feng\affmark[1*] \quad Yixin Gao\affmark[1*] \quad Xin Jin\affmark[2] \quad Runsen Feng\affmark[1] Zhibo Chen \affmark[1]\textsuperscript{,}\affmark[\dag]\\
\affaddr{\small\affmark[1]University of Science and Technology of China\quad}
\affaddr{\small\affmark[2]Eastern Institute of Advanced Study}
}

\maketitle

\let\thefootnote\relax\footnotetext{*~First two authors contributed equally.\\
 \dag~Corresponding author.}

\ificcvfinal\thispagestyle{empty}\fi

\begin{abstract}
    Image compression techniques typically focus on compressing rectangular images for human consumption, however, resulting in transmitting redundant content for downstream applications.
    To overcome this limitation, some previous works propose to semantically structure the bitstream, 
    which can meet specific application requirements by selective transmission and reconstruction.
    Nevertheless, they divide the input image into multiple rectangular regions according to semantics and ignore avoiding information interaction among them, causing waste of bitrate and distorted reconstruction of region boundaries.
    In this paper, we propose to decouple an image into multiple groups with irregular shapes based on a \textbf{customized group mask} and compress them independently.
    Our group mask describes the image at a finer granularity, enabling significant bitrate saving by reducing the transmission of redundant content.
    Moreover, to ensure the fidelity of selective reconstruction, this paper proposes the concept of \textbf{group-independent transform} that maintain the independence among distinct groups.
    And we instantiate it by the proposed \textbf{Group-Independent Swin-Block (GI Swin-Block)}. 
    Experimental results demonstrate that our framework structures the bitstream with negligible cost, and exhibits superior performance on both visual quality and intelligent task supporting. Code is available on \url{https://github.com/Amygyx/GIT-SSIC}.
\end{abstract}

\begin{figure}[htbp]
  \centerline{\includegraphics[width=1.0\linewidth]{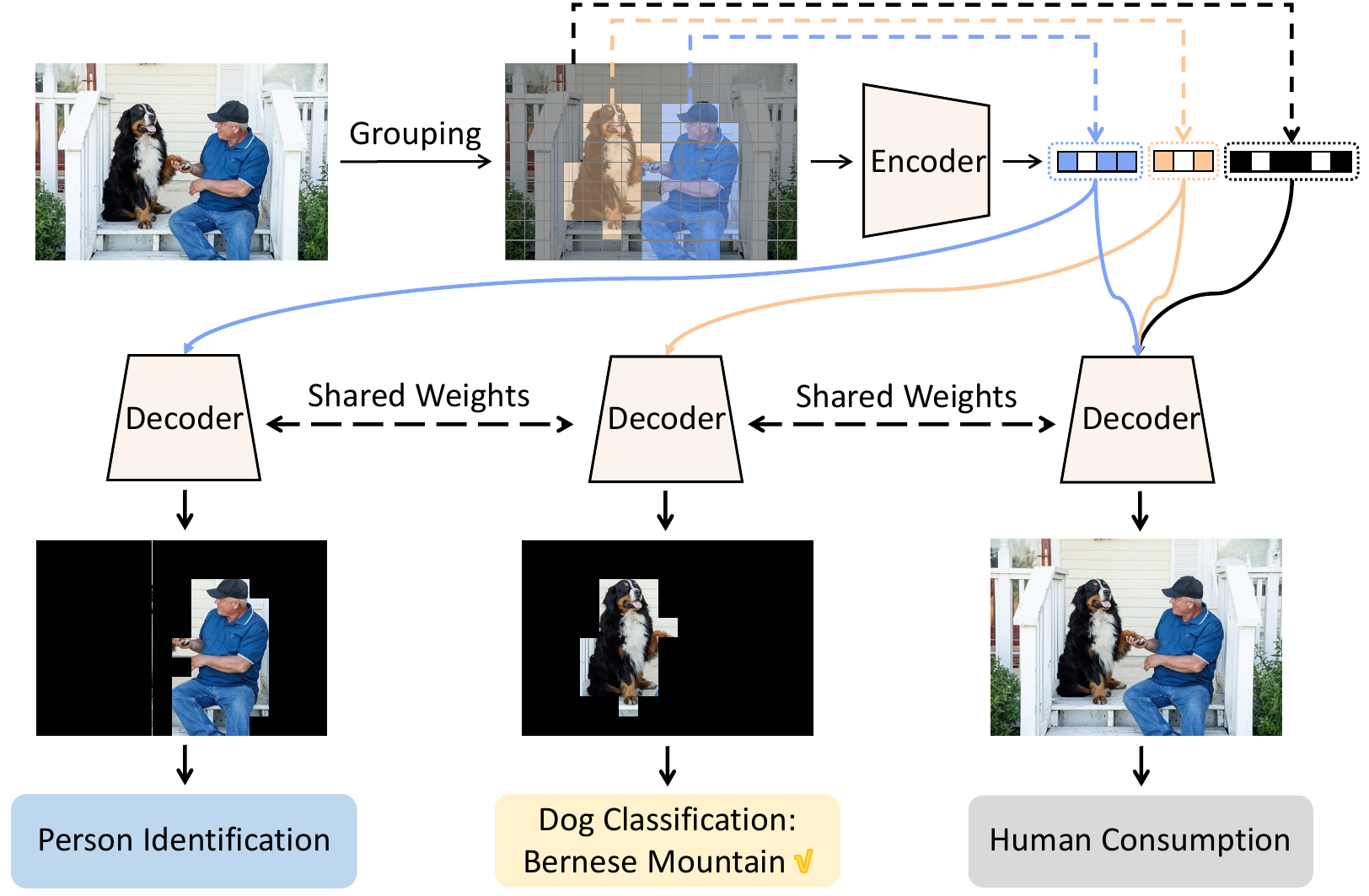}}
    \caption{
    The input image is decoupled into groups according to distinct semantics. Then the semantically structured bitstream (SSB) is generated by compressing the image based on the partitioned groups. The SSB facilitates downstream applications by selective bitstream transmission and partial reconstruction, depending on the specific task requirements.
    }
\label{fig:motivation}
\end{figure}

\section{Introduction}

Image signal serves as a critical information carrier for various applications in modern society.
Image compression techniques aim at converting images into compact representations (\ie bitstreams) to save transmission and storage resources. 
Lossy image compression is one of the most practical techniques, as it allows for the restoration of important content while discarding a small amount of inessential information. 
In the past decades, traditional image compression standards\cite{wallace1992jpeg,rabbani2002overview,wiegand2003overview,sullivan2012overview,bross2021overview} have been extensively studied and utilized.
With the fast development of deep learning, neural image codecs~\cite{balle2017end,balle2018variational,minnen2018joint, cheng2020learned,johnston2018improved,li2018learning,li2020learning,mentzer2018conditional,mentzer2020high,zhu2022transformer,he2022elic} rapidly evolved and achieved promising results. 
Meanwhile, more and more media contents tend to be handled by machine vision algorithms, such as recognition\cite{jia2022visual,he2016deep,liu2021swin,han2021transformer,dosovitskiy2020image}, detection\cite{ren2015faster,redmon2016you, redmon2017yolo9000,lin2017feature,li2022exploring}, and segmentation\cite{he2017mask,liu2018path,bolya2019yolact,long2015fully,badrinarayanan2017segnet,chen2017deeplab,chen2018encoder,xie2021segformer,zheng2021rethinking}.
However, most compression methods are mainly developed for compressing regular rectangular images for human consumption, without considering the efficiency and functionality for downstream tasks or human-machine interaction scenarios.

Recently, the field of image coding for machines (ICM) has emerged to develop a joint efficient and analytical framework for supporting intelligent analytics.
End-to-end optimization of the trade-off between specific task loss and compression rate is a promising way\cite{le2021image,le2021learned,duan2015overview,duan2018compact,chen2019lossy,chen2019toward,singh2020end,ma2018joint,babenko2014neural}, but it lacks generalization for massively diverse applications.
To overcome this limitation, Feng \etal\cite{feng2022image} propose to compress general and compact features learned by self-supervised learning under entropy constraints for supporting downstream tasks. 
Nevertheless, it requires re-training the task model with the proposed features as inputs, which makes the overall performance heavily depend on the effectiveness of feature extraction.
The methods mentioned above are designed specifically for scenarios of compression for machine vision without considering situations in which human involvement is required. Semantically structured image compression (SSIC)\cite{sun2020semantic} proposes to generate a semantically structured bitstream (SSB) by separately compressing rectangular regions of detected objects using a pre-prepared object detection toolbox. Although SSB is efficient in supporting intelligent tasks and human-machine interaction through partial transmission and reconstruction of the bitstream, its division approach based on rectangular regions may encounter problems with overlapped objects. SSIC addresses this issue by replacing the overlapping objects with a larger rectangular region, which can result in a waste of bitrate. In addition, SSIC generates the bitstream of each object by compressing the corresponding latent variables in the latent domain directly, without considering the interaction and dependence of features during the transform process. This can lead to blurry and distorted group boundaries in the partial reconstruction scenario, which in turn affects the reconstruction quality.

Going beyond the rectangular-based division, this paper proposes to decouple the image into multiple groups with irregular shapes based on a customized group mask. 
Then the SSB is generated by independently compressing these groups and can support various requirements via selective transmission and reconstruction.
Notably, the generation of the group mask offers high flexibility in terms of shapes of groups, means of pre-analysis, and partition criteria, enabling customization to suit diverse application scenarios and requirements. 
Moreover, to avoid potential quality degradation in the partial reconstruction scenario, we propose the concept of the group-independent transform, which ensures the independence among groups in the latent representations and therefore the quality of the selective reconstruction will not be affected by the absence of other groups. 
More specifically, we instantiate it by carefully designing the Group-Independent Swin-Block (GI Swin-Block), which is an extension of Swin Transformer\cite{liu2021swin} tailored to our situation and requirements.
GI Swin-Blocks make use of the hierarchical modeling capability of the Swin-Transformer\cite{liu2021swin,zhu2022transformer}, achieving high coding efficiency under the premise of group independence. 
By combining the group mask based decoupling and the Group-Independent Swin-Block, our proposed method can efficiently support various downstream applications including human-machine interaction and machine vision tasks with only one bitstream generated.

The main contributions of our approach are summarized as follows:

\begin{itemize}[label=$\bullet$]
    \item We propose to decouple an image into multiple groups with irregular shapes for structuring the bitstream. Our group mask can decribe spatial division at a finer granularity manner than a typical rectangle, saving bitrate by reducing redundant content transmission.

    \item We propose the group-independent transform and instantiate it by carefully designing the Group-Independent Swin-Block (GI Swin-Block), which maintains powerful transformation capability and ensures the independence among groups in the latent representations.

    \item Experimental results demonstrate that our proposed model achieves state-of-the-art compression ability and superior downstream tasks performance, which is a codec with both high compression efficiency and functionality.

\end{itemize}


\section{Related Works}
\subsection{Image Compression}
\myparagraph{Traditional Image Compression.}
Traditional image compression standards, such as JPEG~\cite{wallace1992jpeg}, JPEG2000~\cite{rabbani2002overview}, HEVC~\cite{sullivan2012overview}, and VVC Intra~\cite{bross2021overview}, have been extensively used in practice after several decades of development. 
These standards rely on transform coding~\cite{goyal2001theoretical}, which decomposes the lossy image compression task into three parts: transform, quantization, and entropy coding.
Each module of these standards is manually designed with multiple modes, and rate-distortion optimization is performed to determine the optimal mode.
However, the completely hand-crafted structure of traditional codecs limits their flexibility and scalability to support various objectives, such as MS-SSIM and classification accuracy, as they cannot be optimized in an end-to-end manner.

\myparagraph{Learned Image Compression.}
In recent years, learned image compression methods based on nonlinear transform coding \cite{balle2020nonlinear} have achieved rapid progress. 
Early works in this area concentrated on enabling end-to-end training by developing differential quantization and rate estimation techniques~\cite{balle2017end,agustsson2017soft,theis2017lossy}. 
Subsequently, a considerable amount of work design powerful neural network modules to enhance the transform~\cite{cheng2020learned,zhu2022transformer}, quantization~\cite{yang2020improving,guo2021soft}, and entropy model~\cite{balle2018variational,minnen2018joint,minnen2020channel,he2022elic}.
As a result, some neural image codecs~\cite{he2022elic,guo2021causal,cheng2020learned,minnen2020channel,minnen2018joint} can achieve comparable or even superior performance to traditional coding standards like HEVC \cite{sullivan2012overview} and VVC \cite{bross2021overview}.
Moreover, some works~\cite{rippel2017real,agustsson2019generative,mentzer2020high} introduce perceptual loss to effectively improve reconstruction quality.
However, existing image compression methods focus on compressing the entire image without considering selective transmission and reconstruction of the bitstream for arbitrary reconstruction, leading to significant bandwidth waste when addressing downstream applications with different requirements.


\subsection{Image Coding for Machines}
Image coding for machines aims at compress source images to serve downstream tasks, such as recognition\cite{jia2022visual,he2016deep,liu2021swin,han2021transformer,dosovitskiy2020image}, detection\cite{ren2015faster,redmon2016you, redmon2017yolo9000,lin2017feature,li2022exploring}, and segmentation\cite{he2017mask,liu2018path,bolya2019yolact,long2015fully,badrinarayanan2017segnet,chen2017deeplab,chen2018encoder,xie2021segformer,zheng2021rethinking}. 
Joint optimization of task loss and bitrate\cite{akbari2019dsslic,hu2020towards,li2021task,le2021image,wang2021towards} in an end-to-end manner is a natural and promising way. Another direction is compress the features corresponding to downstream tasks\cite{duan2015overview,duan2018compact,chen2019lossy,chen2019toward,singh2020end,ma2018joint,babenko2014neural}. 
However, these methods tend to be biased towards specific tasks and may lack generalizability to a wide range of intelligent applications. 
To address the limitation of generalization, a novel representation learning based method is proposed \cite{feng2022image} to learn general and compact features by combining contrastive learning with entropy constraints. 
 The learned features are used to replace the original images as the new source for compression and transmission, resulting in a significant improvement in coding efficiency across various intelligent tasks.
 Nevertheless, this approach necessitates re-training the task model using the proposed features as input, and its overall performance is heavily reliant on the effectiveness of feature extraction. 
 In this paper, we propose a unified compression framework for machine vision support without task-oriented joint training or additional feature learning.
 Moreover, it can also conduct full or selective reconstruction for human perception according to requirements. 

\subsection{Object-oriented Image Compression}
Object-oriented image compression is first proposed in Mpeg-4 Visual\cite{sikora1997mpeg,katsaggelos1998mpeg,ebrahimi2000mpeg} by compressing visual object planes (VOP) with arbitrary shapes, targeting content-based interactivity and scalability for the image compression technique. 
However, due to the difficulty in annotating the complex and intricate shapes of the target and the lack of actual corresponding downstream applications, MPEG-4 Visual has not been widely used.
Recently, with the fast development of deep learning\cite{he2016deep,vaswani2017attention}, the practical scenarios of image compression for downstream applications have increased rapidly. 
Sun \etal\cite{sun2020semantic,jin2022semantically} propose the semantically structured image compression (SSIC) of decoupling the image into objects and compressing the corresponding latent variables independently of the neural image codec to generate the semantically structured bitstream. 
Nevertheless, the rectangular-based division adopted by SSIC lacks generalization to irregular objects and flexibility for customization. 
Additionally, selecting the representation of objects directly in the latent space ignores the interaction among elements during the encoding transform process, leading to significant edge distortion of objects in partial reconstruction. 
This paper proposes to conduct semantically structured image compression by decoupling the image into irregular groups based on the group mask, which is flexible and customizable. 
Furthermore, we propose the group-independent transform conducted during both the encoding and decoding processes, which enables efficient partial or complete reconstruction of images under various requirements.

\section{Method}
\subsection{Overview}

\begin{figure*}[htbp]
\centerline{\includegraphics[width=1.0\linewidth]{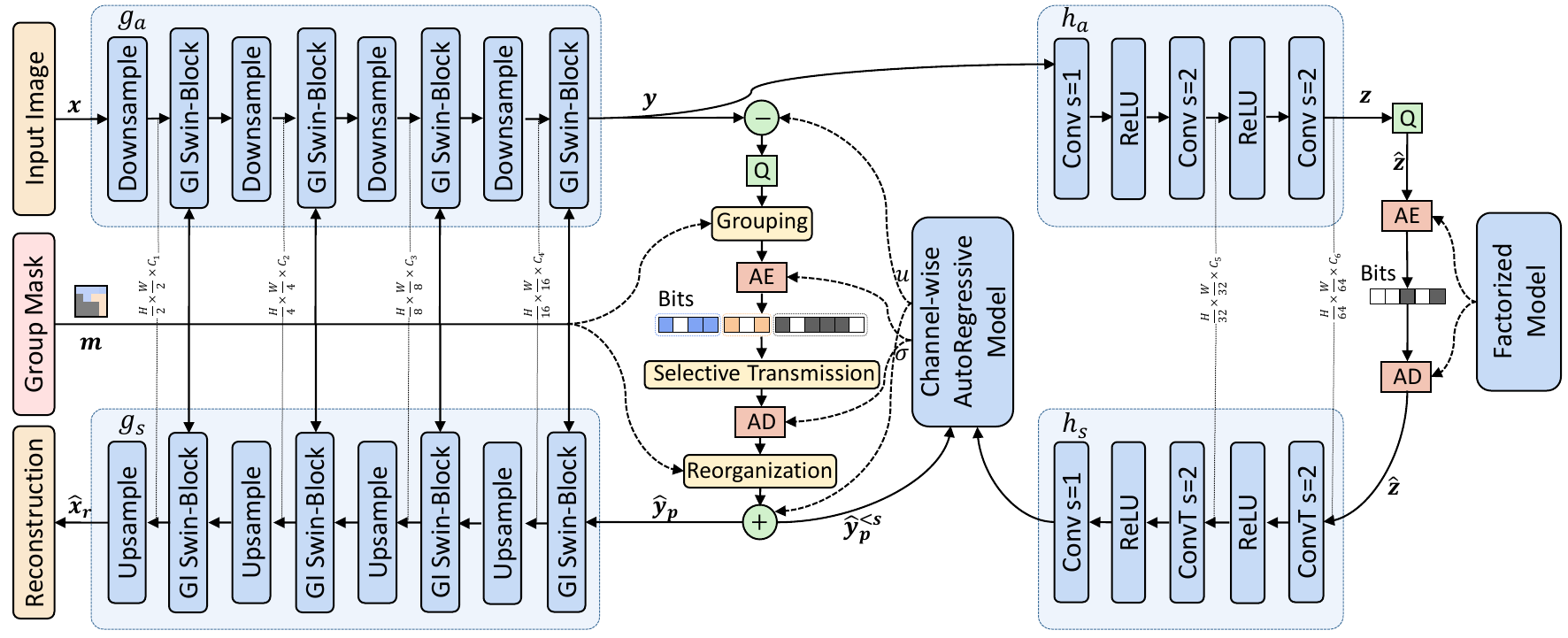}}
    \caption{The network architecture of our proposed model with the channel-wise auto-regressive model (ChARM). ConvT denotes transposed convolution. AE and AD are respectively arithmetic encoding and arithmetic decoding. In Ours-Hyper model, we remove the ChARM component and instead output $\mu$ and $\sigma$ directly from the hyperdecoder $h_s$.}
\label{fig:codec_architecture}
\end{figure*}

The proposed method is an efficient and flexible version of semantically structured image compression~\cite{sun2020semantic}.
It can satisfy multiple application requirements via transmitting and reconstructing partial spatial regions guided by a block-wise group mask.
Our model takes the group mask as guidance to ensure that interactions of transform only occur within the same group, thereby achieving group independence during the redundancy removal process.
The group mask, which is generated based on the pre-analysis, such as object detection, semantic segmentation, and saliency detection, provides high flexibility and customization for structuring the bitstream and is considered as side information. 
Then the entropy coding process is conducted on the variables of each group distinctly, resulting in a bitstream structured by semantics. 
The bitstream can be partially or fully transmitted according to the requirements of the decoder side, and then the receiver conducts entropy decoding on the bitstream and reorganizes the spatial-wise arrangement of latent variables according to the group mask and the group indexes. 


The overall network structure is illustrated in Fig.~\ref{fig:codec_architecture}, which incorporates the group mask and GI Swin-Block for guiding and instantiating the group-independent transform, respectively.
Section \ref{sec: Methods:group mask} and Section \ref{sec: Methods:group independent transform} introduce the group mask and group independent, respectively.
The detailed implementation of our network architecture is presented in Section \ref{sec: Methods:networkstructure}.

\subsection{Group Mask}
\label{sec: Methods:group mask}
\begin{figure}[t]
\centerline{\includegraphics[width=1.0\linewidth]{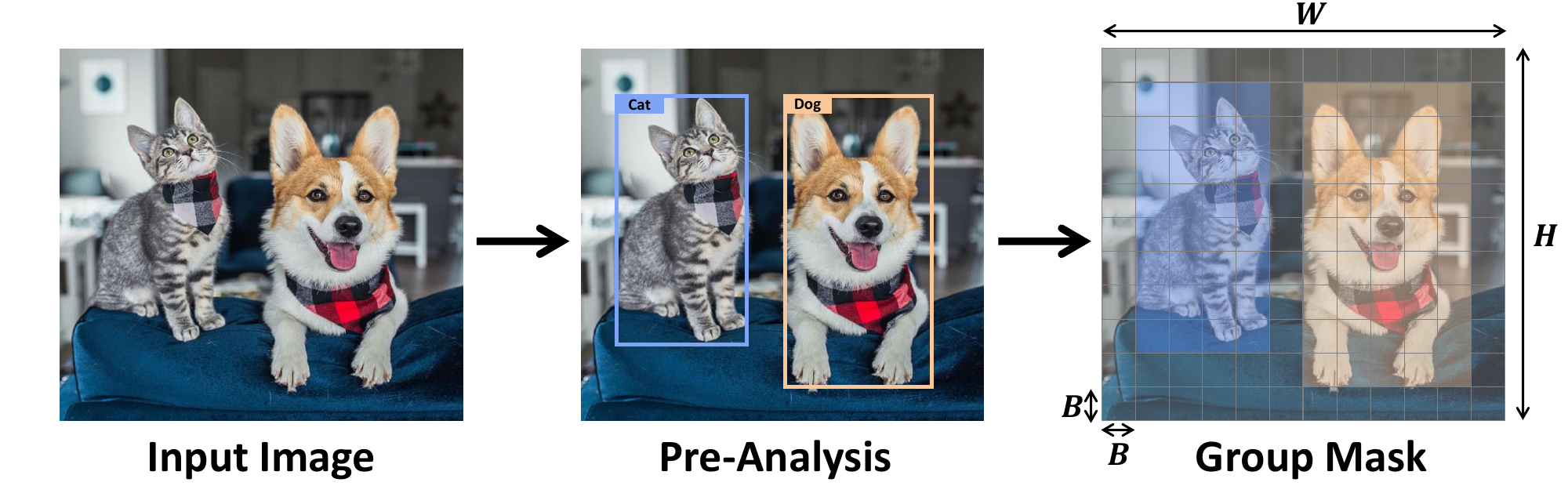}}
    \caption{
    An example of group mask generation. The input image is pre-analyzed by object detection, then the group mask is generated based on the results of the pre-analysis.
    }
\label{fig:illustration of group mask}
\end{figure}

Before compression, the group mask is generated according to the results of pre-analysis techniques such as object detection, instance segmentation, and saliency detection. Figure \ref{fig:illustration of group mask} provides an example of this process.
More specifically, for the input image $x$ with a height of $H$ and a width of $W$, the spatial resolution of the group mask is the same as the input image, and it consists of $\frac{H}{B} \times \frac{W}{B}$ blocks, where $B$ is the side length of each block.
And the value of it indicates the index of the corresponding group.

\begin{figure}[t]
\centerline{\includegraphics[width=1.0\linewidth]{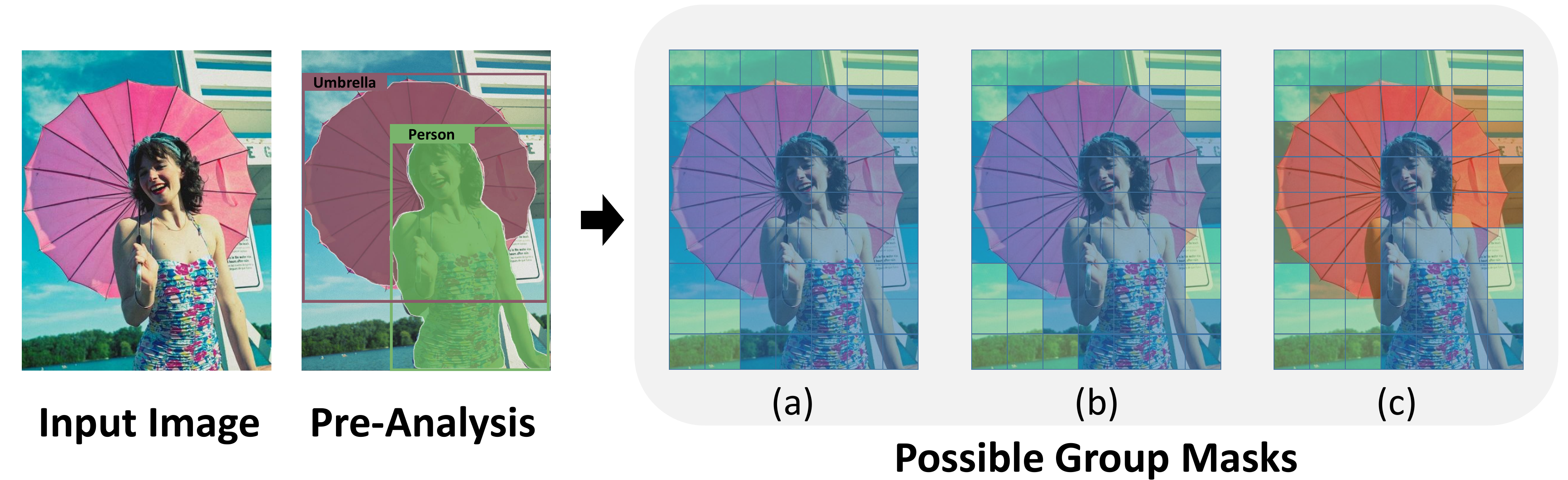}}
    \caption{
    Illustration of flexibility to customize the group mask: (a). Taking overlapping objects as one group with the guidance of bounding boxes. (b). Taking overlapping objects as one group with the guidance of instance masks. (c) Taking overlapping objects as distinct groups with the guidance of instance masks. 
    }
\label{fig:flexibility of group mask}
\end{figure}

Our proposed group mask can clearly divide the elements in the latent variable space after the one-pass downsampling transform. 
Besides, downsampling the group mask by a factor of $B$ before conducting compression can greatly reduce the overhead bitrate. 
The block size and the overhead bitrate have an inverse relationship. 
Furthermore, as shown in Figure \ref{fig:flexibility of group mask}, the generation of the group mask can be flexibly customized, in terms of the way of pre-analysis and the criteria of block allocation, \etc. 

\subsection{Group-Independent transform}
\label{sec: Methods:group independent transform}

Lossy image compression based on transform coding can be divided into three modularized components: transform, quantization, and entropy coding. 
While quantization and entropy coding do not affect the generation and usage of the semantically structured bitstream (SSB), traditional transform applied to the entire image to remove spatial redundancy inevitably creates inter-group dependencies during compression. 
In situations of selective transmission and reconstruction, the incomplete representation that results from the reorganization of a subset of all groups can lead to inaccurate reconstruction due to the lack of inter-group dependencies.
Therefore, we propose the concept of the group-independent transform for semantically structured bitstream generation. 
The basic idea is to constrain the transform to be conducted only within each group. 

\begin{figure}[t]
  \centering
  \vspace{-2mm}
  \begin{subfigure}[b]{0.48\textwidth}
    \includegraphics[width=\textwidth]{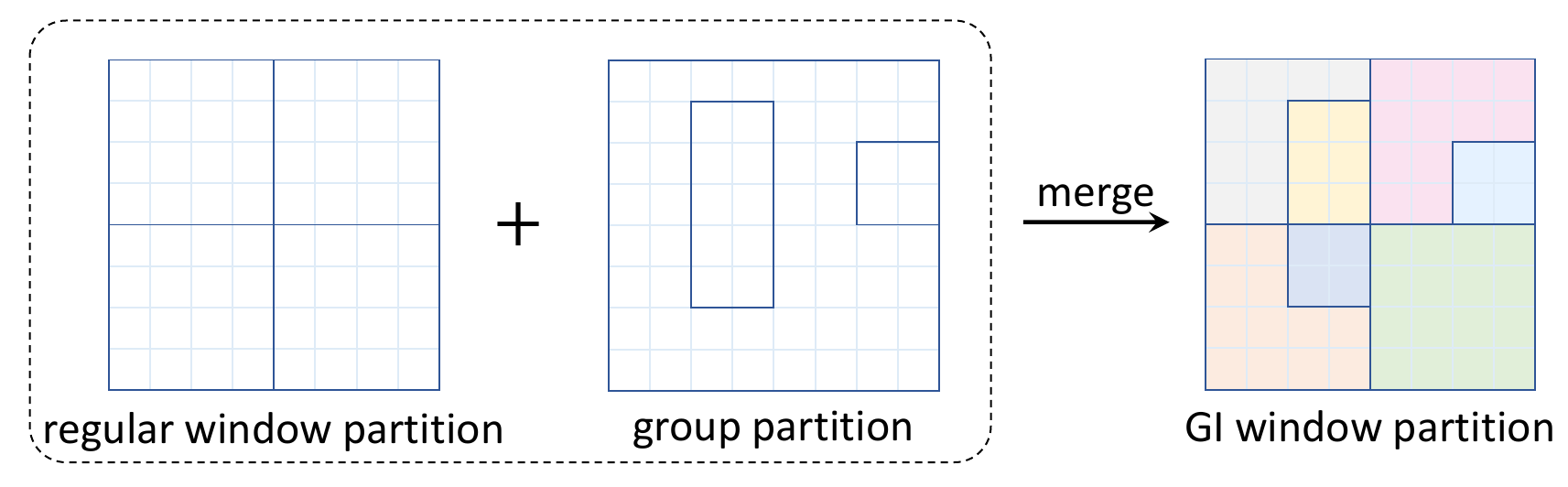}
    \caption{GI window partition under regular window partition.}
    \label{fig:GI_swin_block_regular}
  \end{subfigure}
  \begin{subfigure}[b]{0.48\textwidth}
    \includegraphics[width=\textwidth]{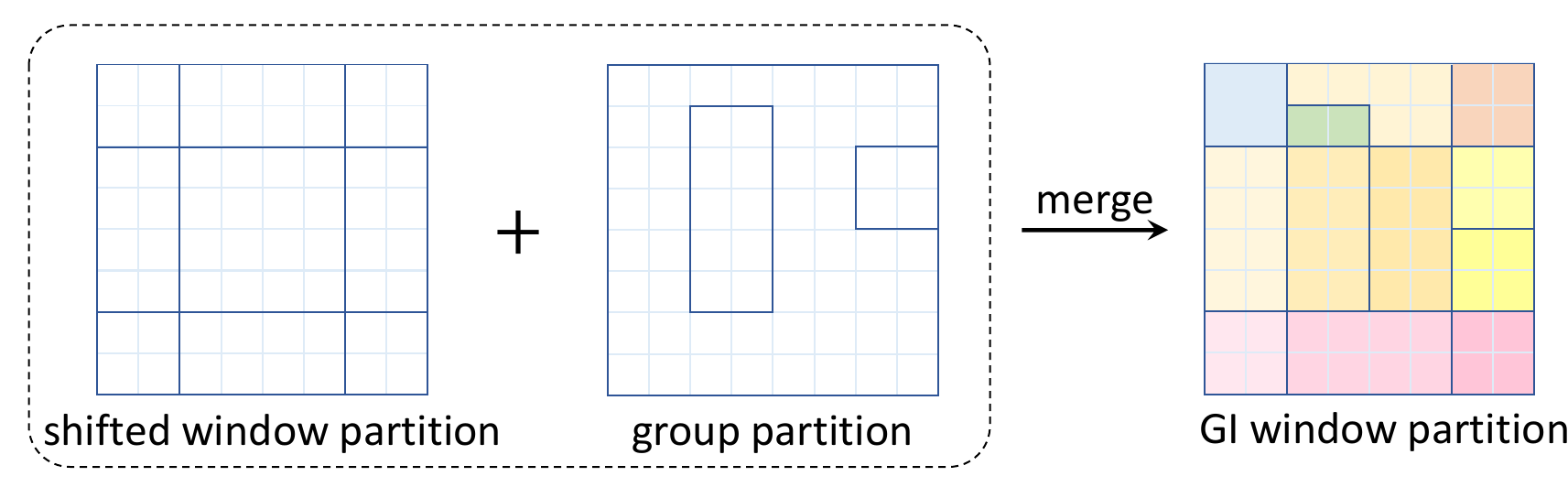}
    \caption{GI window partition under shifted window partition.}
    \label{fig:GI_swin_block_shift}
  \end{subfigure}
  \caption{Group-independent window partition of GI Swin-Block under regular and shifted window partition. Self-attention is conducted inside each local region.}
  \label{fig:GI_swin_block}
\vspace{-2mm}
  
\end{figure}

A straight and natural way is by using transformer \cite{vaswani2017attention} with the customized attention map corresponding to the group mask. 
However, image compression differs significantly from high-level understanding tasks, which the transformer excels at. 
The presence of numerous complex long-range dependencies can impede convergence and adversely affect the final performance. 
In order to achieve high coding efficiency while maintaining the group-independent property, and inspired by the strong hierarchical representation modeling capability of Swin-Transformer\cite{liu2021swin,zhu2022transformer}, we extend the Swin-Block of Swin-Transformer to our proposed group-independent Swin-Block (GI Swin-Block). 
The GI Swin-Block serves as the core component of the transform.
To be more specific, as shown in Figure \ref{fig:GI_swin_block_regular}, by merging the window partition and the group partition into the group-independent window partition, allowing the self-attention to be computed within the partitioned local regions, enabling us to achieve high coding efficiency while maintaining the group-independent property. 
Additionally, the cross-window connections are introduced by merging the shifted window partition and group partition similarly, as depicted in Figure \ref{fig:GI_swin_block_shift}.


\subsection{Network architecture and pipeline}
\label{sec: Methods:networkstructure}

Figure~\ref{fig:codec_architecture} illustrate the proposed architecture. 
The main transform is performed in a staggered manner through the incorporation of upsample/downsample and group-independent Swin-Block (GI Swin-Block).
Upsample contains a pixel shuffle operation and a convolution layer with $1\times 1$ kernel size, whereas downsample contains a pixel unshuffle operation and a convolution layer with $1\times 1$ kernel size.
For entropy coding, we adopt the Mean \& Scale (M\&S) Hyperprior model~\cite{minnen2018joint} with the strong channel-wise auto-regressive model (ChARM)~\cite{minnen2020channel} to predict the probability distribution of latent variables.

Specifically, given a source image $\boldsymbol x$, the encoder $g_a$ converts it to latent representations $\boldsymbol y$ conditioned on the given group mask $\boldsymbol m = \sum_{i=1}^{N} \boldsymbol{m}_{i}$, where $ i\in\{1, 2, 3, \ldots, N\}$, indicating the index of the corresponding group in all $N$ groups. 
In each $\boldsymbol{m}_{i}$, the values of positions corresponding to the $i_{th}$ group are set as $i$, and all values of other positions are set as 0.
Similarly, we can define $\boldsymbol x = \sum_{i=1}^{N} \boldsymbol{x}_{i} $ and $\boldsymbol y = \sum_{i=1}^{N} \boldsymbol{y}_{i}  $ from the perspective of group, and transform only occurs within each group, which is given by
\vspace{-1mm}
\begin{equation}
    \boldsymbol{y}_{i}  = g_{a}\left ( \boldsymbol{x}_{i} | \boldsymbol{m}_{i} \right ).
    \label{eq:encoding transform}
\end{equation}
\vspace{-1mm}
Thanks to the group-independent transform, we can easily implement the transform of all groups within one forward-pass, thus, the process can also be written as $\boldsymbol{y}  = g_{a}\left ( \boldsymbol{x} | \boldsymbol{m} \right )$.

Then the latents are quantized to discrete representations $\boldsymbol {\hat{y}}$ and selectively transmitted to the decoder side with predicted probability distribution through the entropy model. 
Concretely, the side information $\boldsymbol z$ is extracted by the hyper-encoder $\boldsymbol z = h_{a} \left ( \boldsymbol y \right )$. The quantized hyper-latent $\boldsymbol {\hat{z}}=Q \left ( \boldsymbol z \right )$ is modeled and entropy-coded with a learned factorized prior. 
In ChARM, latent $\boldsymbol{\hat{y}}$ is split into $S$ (we choose $S$=10) slices along with channel dimension, and each slice $\boldsymbol{\hat{y}}_s$ is entropy-coded based on the previous slices $\boldsymbol{\hat{y}}_{<s}$. $\boldsymbol{\hat{y}}$ is modeled by a conditional Gaussian distribution convolving with a unit
uniform noise.

After selectively entropy decoding the quantized latents, we have $\boldsymbol{\hat{y}}_{p} = \sum_{j \in \mathbf{P}} \boldsymbol{\hat{y}}_{j}$, where $\mathbf{P} \subseteq \left \{ 1, 2, \ldots  ,N \right \} $ is the gather of required groups. The reconstruction image 
$\boldsymbol{\hat{x}}_{p} = \sum_{j \in \mathbf{P}} \boldsymbol{\hat{x}}_{j} $ will be calculated as follows
\vspace{-1mm}
\begin{equation}
    \boldsymbol{\hat{x}}_{j}  = g_{s}\left ( \boldsymbol{\hat{y}}_{j} | \boldsymbol{m}_{j} \right ).
    \label{eq:decoding transform}
\end{equation}
Similar to Equation (\ref{eq:encoding transform}), Equation (\ref{eq:decoding transform}) can also be written as $\boldsymbol{\hat{x}}  = g_{s}\left ( \boldsymbol{\hat{y}} | \boldsymbol{m} \right )$.
Note that the rounding operation is non-differentiable, thus we apply uniform noise for learning the entropy model and use a rounded one as the input of the decoder $g_{s}$ as in~\cite{minnen2020channel}.

\section{Experiments}

\begin{figure*}[htbp]
    \vspace{-3mm}
    \centering
    \begin{subfigure}[b]{0.48\textwidth}
        \captionsetup{skip=-1pt}
        \centerline{\includegraphics[width=0.87\textwidth]{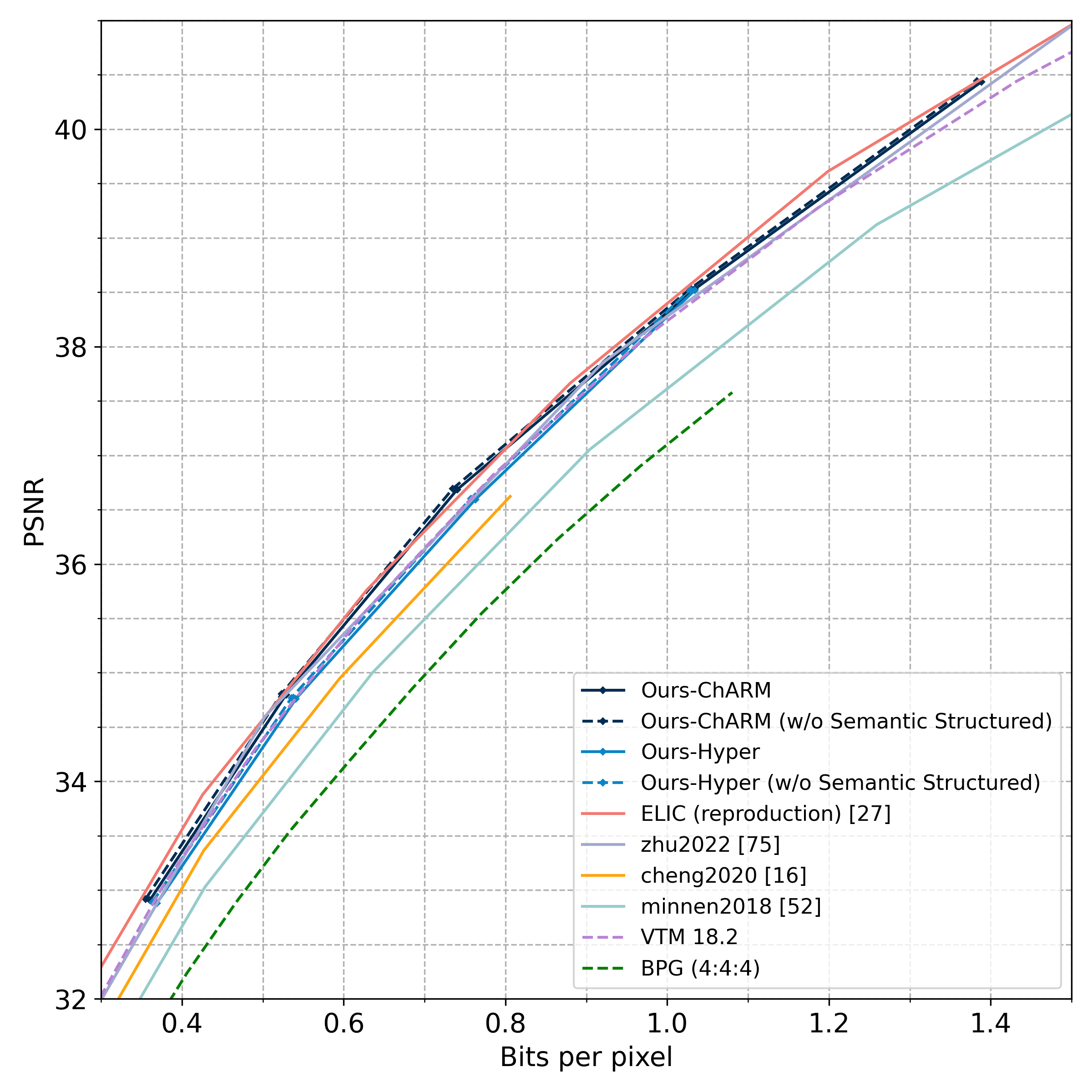}}
        \caption{Entire image reconstruction on Kodak.}
        \label{fig:RD_recon_full}
    \end{subfigure}
    \begin{subfigure}[b]{0.4\textwidth}
        \begin{subfigure}[b]{1.0\textwidth}
            \setlength{\abovecaptionskip}{-2pt}
            \includegraphics[width=1.0\textwidth]{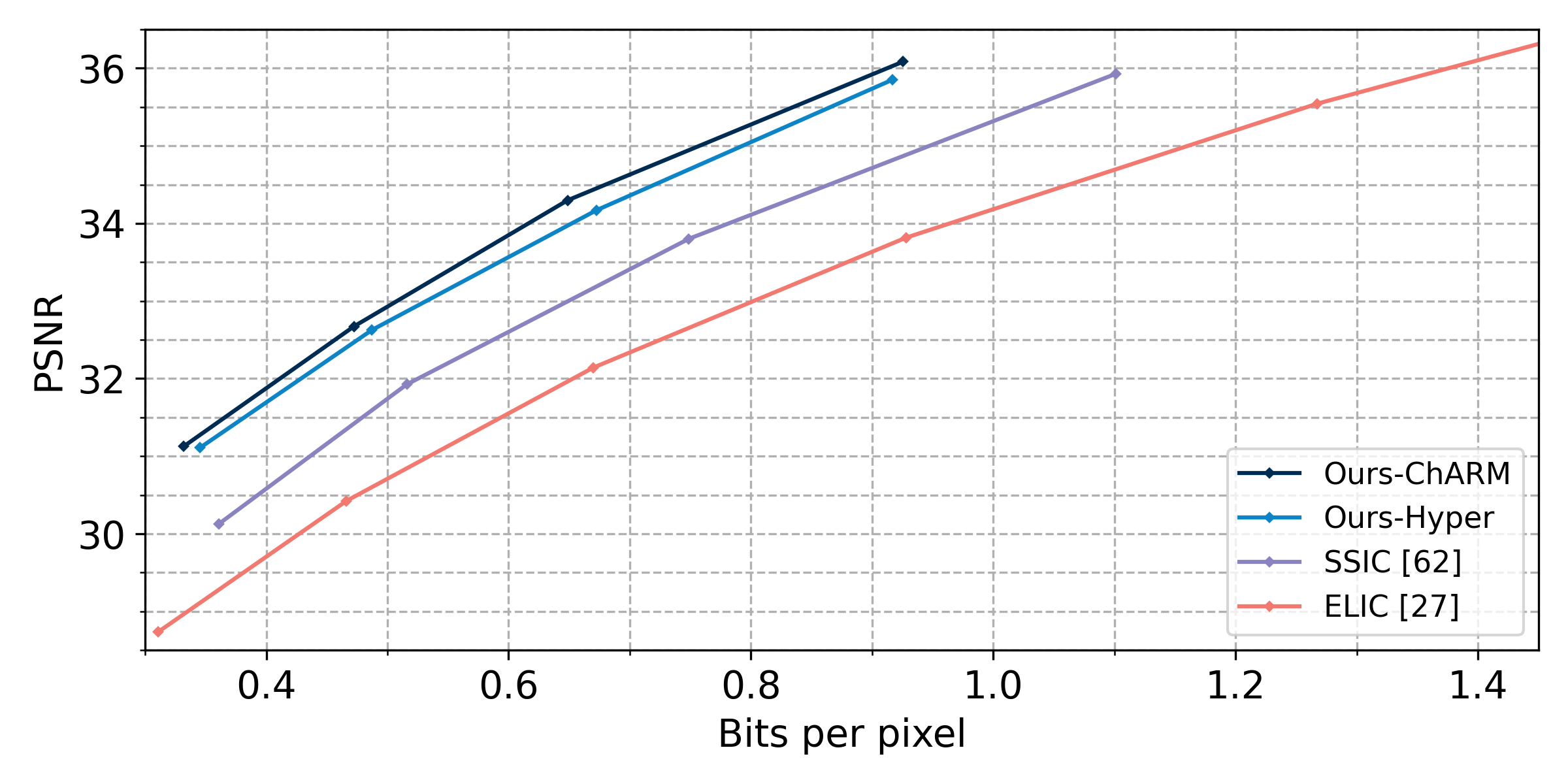}
            \caption{All categories objects reconstruction on COCO.}
            \label{fig:RD_recon_objects}
            \includegraphics[width=1.0\textwidth]{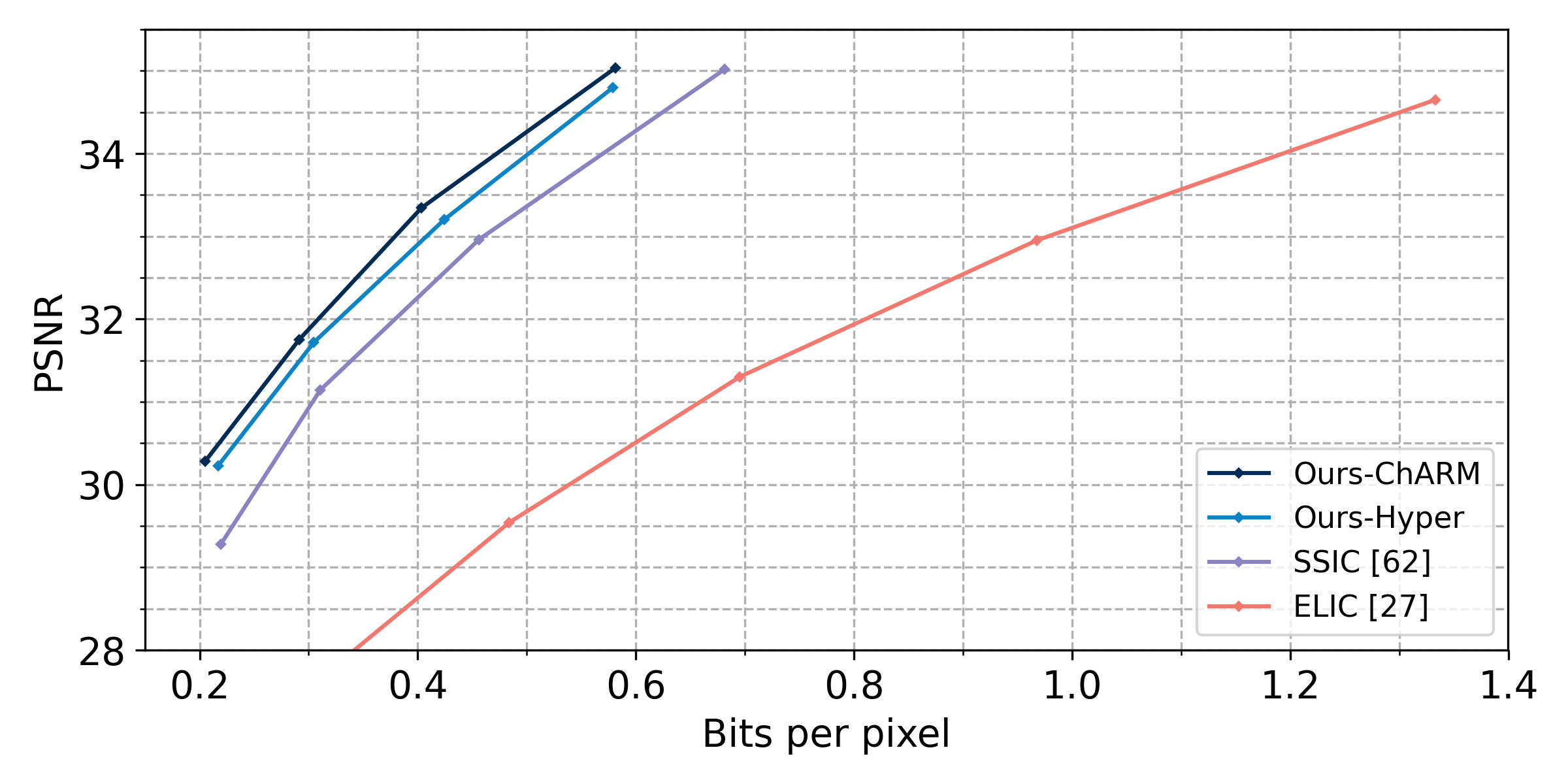}
            \caption{Human category objects reconstruction on COCO.}
            \label{fig:RD_recon_human}
        \end{subfigure}
    \end{subfigure}
    \vspace{-2mm}
    \caption{Rate distortion comparison of compression efficiency on both entire and partial reconstruction scenarios. The PSNR of partial reconstructions is only calculated on the pre-detected bounding boxes.}
    \label{fig:main}
    \vspace{-4mm}
    
\end{figure*}

\begin{figure}[htbp]
  \centering
  \setlength{\abovecaptionskip}{0mm}
  \setlength{\belowcaptionskip}{-5pt}
  \begin{subfigure}[b]{0.23\textwidth}
    \includegraphics[width=\textwidth]{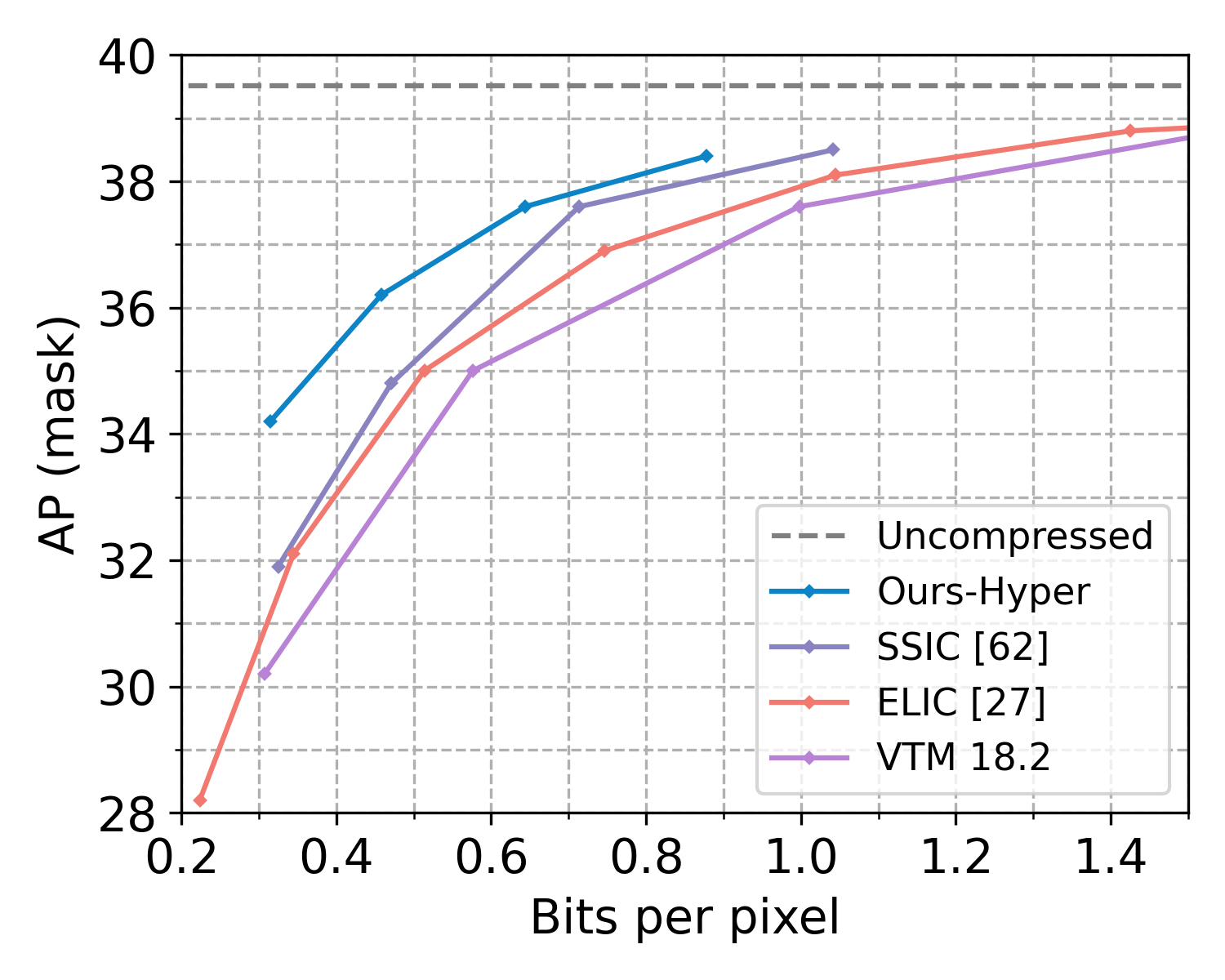}
  \end{subfigure}
  \hspace{-1mm}
  \begin{subfigure}[b]{0.23\textwidth}
    \includegraphics[width=\textwidth]{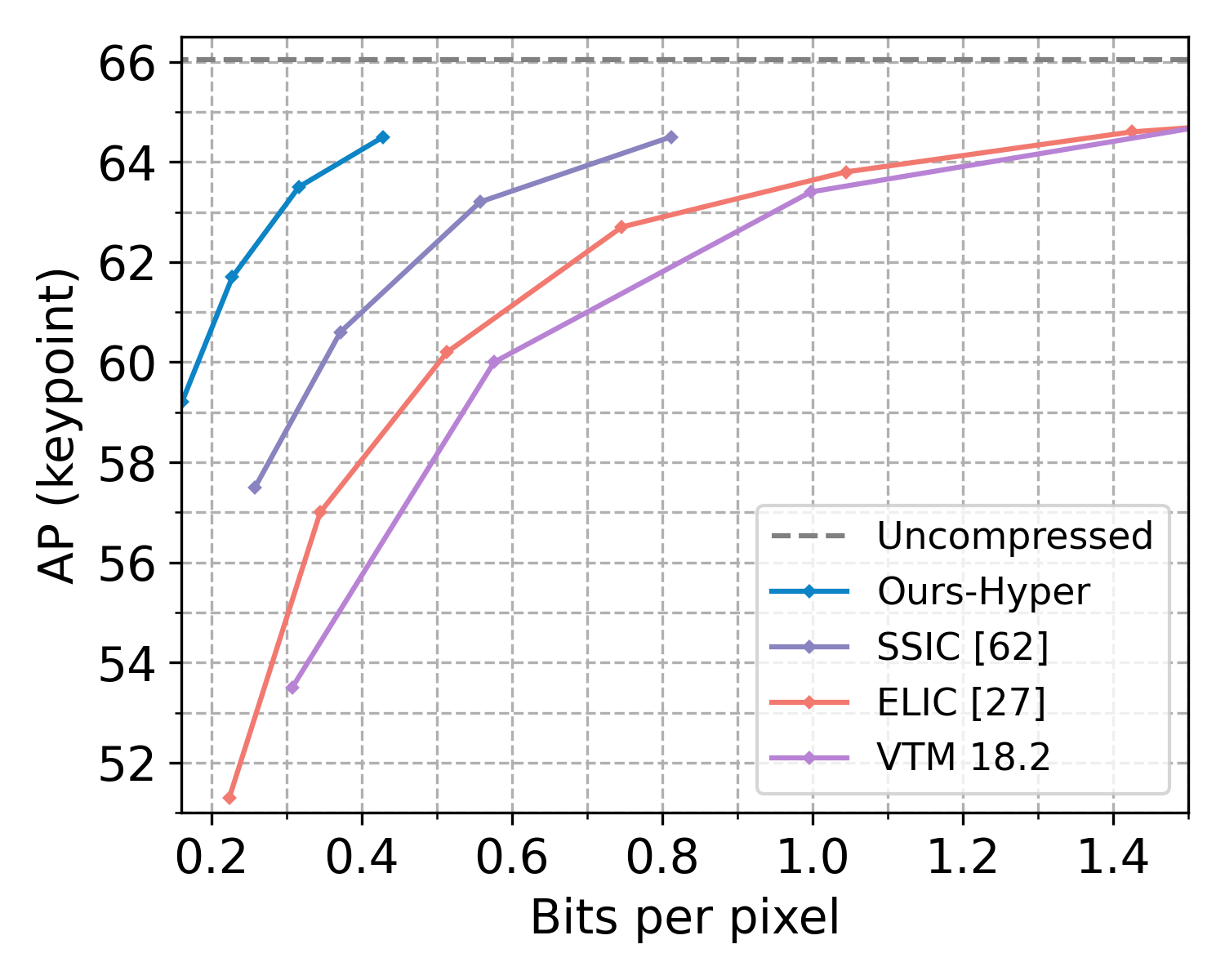}
  \end{subfigure}
  \caption{Performance comparison on instance segmentation (left) and pose estimation (right) on COCO.}
  \label{fig:DownstreamTasks}
    \vspace{-3mm}
  
\end{figure}

\subsection{Experiment Setup}
\myparagraph{Image Codec.}
We train our high bitrate models ($>=0.6$bpp) on the Flickr 2K dataset, which has the same setting as \cite{zhou2019end} and train our low bitrate models ($<0.6$bpp) on the COCO 2017 training set\cite{lin2014microsoft}. 
Ours-Hyper are trained with 3.2M iterations, while Ours-ChARM models are trained by initializing their weights from higher bitrate hyperprior models for another 1.2M iterations.
Each batch contains 8 random $256\times256$ crops from the training dataset. 
The learning rate is set as $5e-5$ and is decayed by a factor of $10$ at $2.8M$ iterations. 
Training loss $L=R+\beta D$ is the weighted combination of the rate-distortion trade-off, balanced by the Lagrange multiplier $\beta$. 
We adopt the mean squared error (MSE) in the RGB color space as the distortion metric $D$.
The bitrate range of our image codecs is achieved by applying different $\beta$ that $\beta\in\{512, 1024, 2048, 4096\}$. 
During the training stage, we randomly generate the group masks for the training of the group-independent transform, the detailed operations are shown in the supplementary material. 

\myparagraph{Evaluation Datasets and Protocol.}
For the entire image reconstruction quality, we use the widely-used Kodak dataset \cite{kodak} to evaluate the coding efficiency of our models.
We evaluate the selective reconstruction quality on objects of all categories and human objects using 40 and 20 images, which are randomly selected from the COCO 2017 validation set.
The PSNR of the region of interest serves as the metric for measuring the objective quality. Additionally, the bits per pixel (bpp) are calculated by dividing the total bits of the transmitted bitstream by the number of pixels in the region of interest.
To verify the effectiveness of our proposed method in supporting downstream tasks, we conduct experiments on instance segmentation and pose estimation using the COCO 2017 validation set. \cite{lin2014microsoft}. 
This dataset is widely used for dense prediction tasks and comprises 118K training, 5K validation, and 20K test-dev images. 
We evaluate the performances of instance segmentation and human keypoint detection using Mask R-CNN (X101-FPN backbone) and Keypoint R-CNN (X101-FPN backbone), respectively, implemented in the Detectron2 toolbox \cite{wu2019detectron2}. 

\begin{figure*}[t]
\vspace{-3mm}
\centerline{\includegraphics[width=1.0\linewidth]{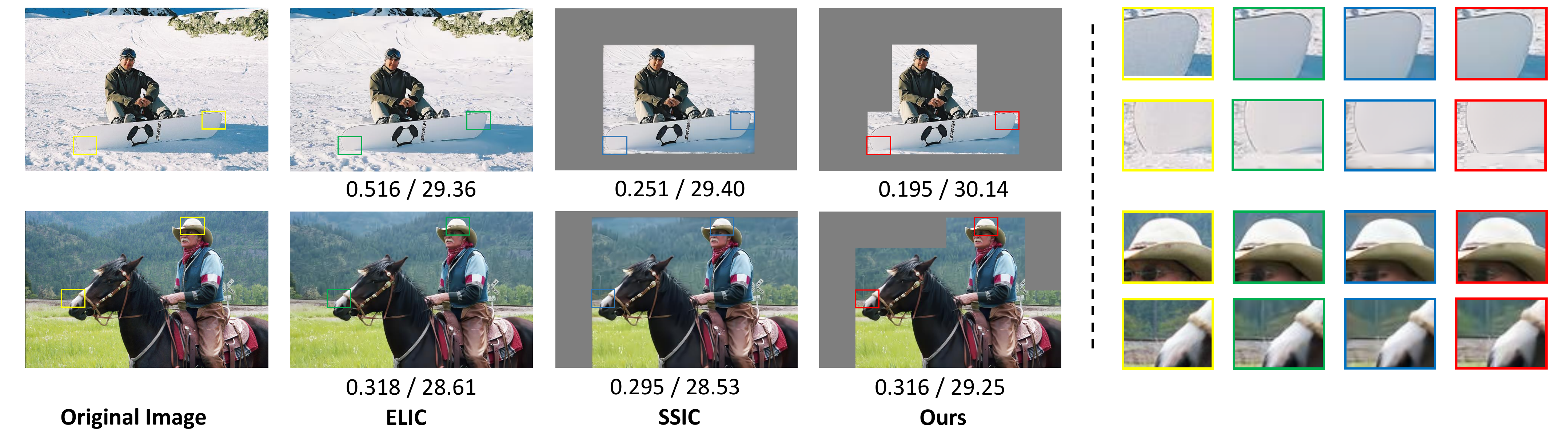}}
    \vspace{-2.5mm}

    \caption{Visualization of reconstructed images for human consumption from the COCO 2017 dataset, in the scenario where foreground objects are the regions of interest. The numbers below the images correspond to bits per pixel and PSNR.}
    
\label{fig:visualize_comparison}
\end{figure*}

\vspace{-2.5mm}

\begin{figure*}[htbp]
\centerline{\includegraphics[width=1.0\linewidth]{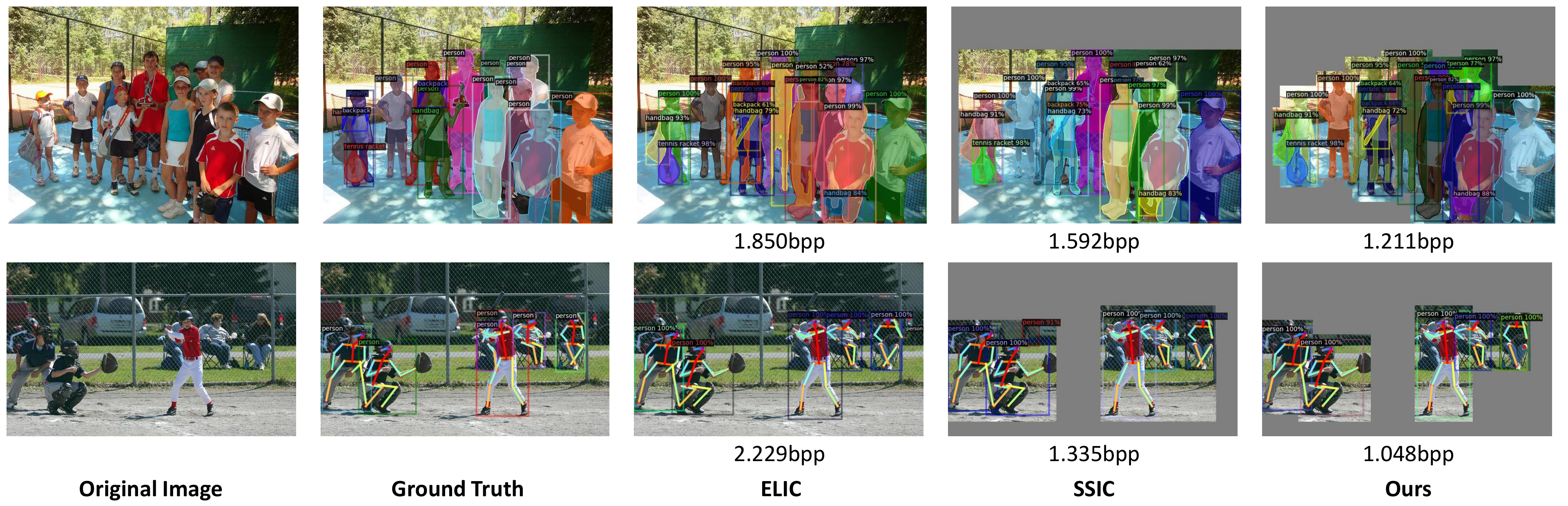}}
    \vspace{-2.5mm}

    \caption{Visualization of reconstructed images for downstream tasks from the COCO 2017 dataset. The first line corresponds to instance segmentation, and the second one corresponds to pose estimation.}
\label{fig:visualize_downstream_task}
\vspace{-5.5mm}
\end{figure*}

\subsection{Quantitative Results}

For comparisons, we choose recent neural image codecs~\cite{minnen2018joint,cheng2020learned,zhu2022transformer,he2022elic}, powerful classical image codecs~\cite{sullivan2012overview,bross2021overview}, and the functional codec that supports semantically structured bitstream~\cite{sun2020semantic}.
Among them, ELIC\cite{he2022elic} is reproduced and performs closely as their report.
It is critical to emphasize that the encoder does not possess prior knowledge regarding the regions of interest required by the decoder. 
And we utilize object detection as the pre-analysis by default, which is one of the most fundamental visual analysis tools.
The semantically structured bitstreams are generated based on the bounding boxes detected by the detector of Mask R-CNN (X101-FPN backbone), and are selectively decoded and reconstructed into images for evaluation according to the requirements of relative semantics.
In all experiments of quantitative results, the block size $B$ of group masks is set to 32 to balance the trade-off between performance and overhead bitrate, and we take the union of overlapping groups as one group. More experiments compared with other methods of image coding for machines are presented in the supplementary.

\myparagraph{Reconstruction Quality.}
For entire image reconstruction, Figure.~\ref{fig:RD_recon_full} demonstrates that our proposed model achieves state-of-the-art rate-distortion performance, surpassing VTM 18.2 in terms of PSNR at all bitrates. 
Notably, structuring the bitstream according to semantics introduces negligible effects. 
Therefore, our model has a strong potential to achieve both high coding efficiency and functionality simultaneously. 
For partial image reconstruction in scenarios of specific interest, as shown in Figure.~\ref{fig:RD_recon_objects} and Figure.~\ref{fig:RD_recon_human}, our model has achieved a significant improvement compared to other codecs. 
Specifically, the semantically structured bitstream enabled our model and SSIC~\cite{sun2020semantic} to avoid transmitting and decoding bitstreams corresponding to the entire image, which is required of ELIC~\cite{he2022elic}.
Additionally, our model's flexible block-level group mask and group-independent transform spatially decouple latents from the semantic level, resulting in sharper and more realistic boundaries compared to SSIC~\cite{sun2020semantic}. Moreover, in the case of overlapping regions, our model can significantly save bitrate by grouping them into an irregular group instead of replacing them with a larger bounding box, which would introduce more distortion.

\myparagraph{Downstream Task Supporting.}
Figure.~\ref{fig:DownstreamTasks} demonstrates our excellent performance on instance segmentation and pose estimation. On instance segmentation, our model achieves significantly improved performance at low bitrates ($<$0.6bpp) compared with other methods, which can be attributed to the faithful reconstruction of RoI boundaries in our approach.
It is worth noting that the superior performance of our method on pose estimation is due to the ability of our model to preserve fine details and boundaries, which are critical for accurately localizing human keypoints. 
In addition, human objects are a minority in the image and are often sparsely distributed, and compared with SSIC, our method can significantly reduce bitrate by avoiding transmitting the latents corresponding to the rectangular area containing all overlapped objects. 

\vspace{-1.2mm}

\subsection{Qualitative Results}
\vspace{-1.2mm}

\myparagraph{Reconstruction for Human Consumption.} 
When the reconstructed images are mainly intended for human consumption with specific semantic interests, as shown in Figure \ref{fig:visualize_comparison}, bitstreams generated by codecs designed for compressing regular rectangular images are required to be fully transmitted and serve full reconstruction, regardless of the requirements and the content of images.
Although SSIC \cite{sun2020semantic} partially addresses this issue through generating SSB through bounding box based division, it lacks flexibility and may introduce additional irrelevant disturbance when merging overlapping objects into larger bounding boxes. 
Moreover, SSIC generates the bitstream by directly dividing the corresponding positions in the latent variable space after the transform based on ConvNets, which can lead to lost dependencies during selective reconstruction, resulting in blurred and distorted region boundaries.
With group mask based partitioning and group-independent transform, our method decouples images to generate the SSB more efficiently while ensuring that no distortion or blurring is introduced after selective reconstruction, leading to both significant bitrate saving and a better visual experience.


\myparagraph{Reconstruction for Intelligent Tasks.}
When using reconstructed images for downstream intelligent tasks, selectively transmitting and decoding the semantically structured bitstream based on the prior knowledge of relative semantics can significantly save bitrate. 
As illustrated in Figure \ref{fig:visualize_downstream_task}, conventional codecs designed for compressing rectangular images require the entire bitstream to be transmitted and decoded. 
Downstream models then perform intelligent analytics on the fully reconstructed image, resulting in a waste of bitrate. 
Similar to image reconstruction for human consumption, SSIC can perform selective transmission to save bitrate. However, its bounding box based partitioning may not be optimal for bitrate savings, and the resulting blurred and distorted region boundaries and irrelevant content can further impede downstream intelligent analytics.
(misidentifying the car as a person shown in Figure \ref{fig:visualize_downstream_task}).
Our method can more efficiently generate the SSB and support various downstream tasks with specific requirements with both accuracy and coding efficiency. 

\begin{figure}[t]
\vspace{-3mm}
\centerline{\includegraphics[width=1.0\linewidth]{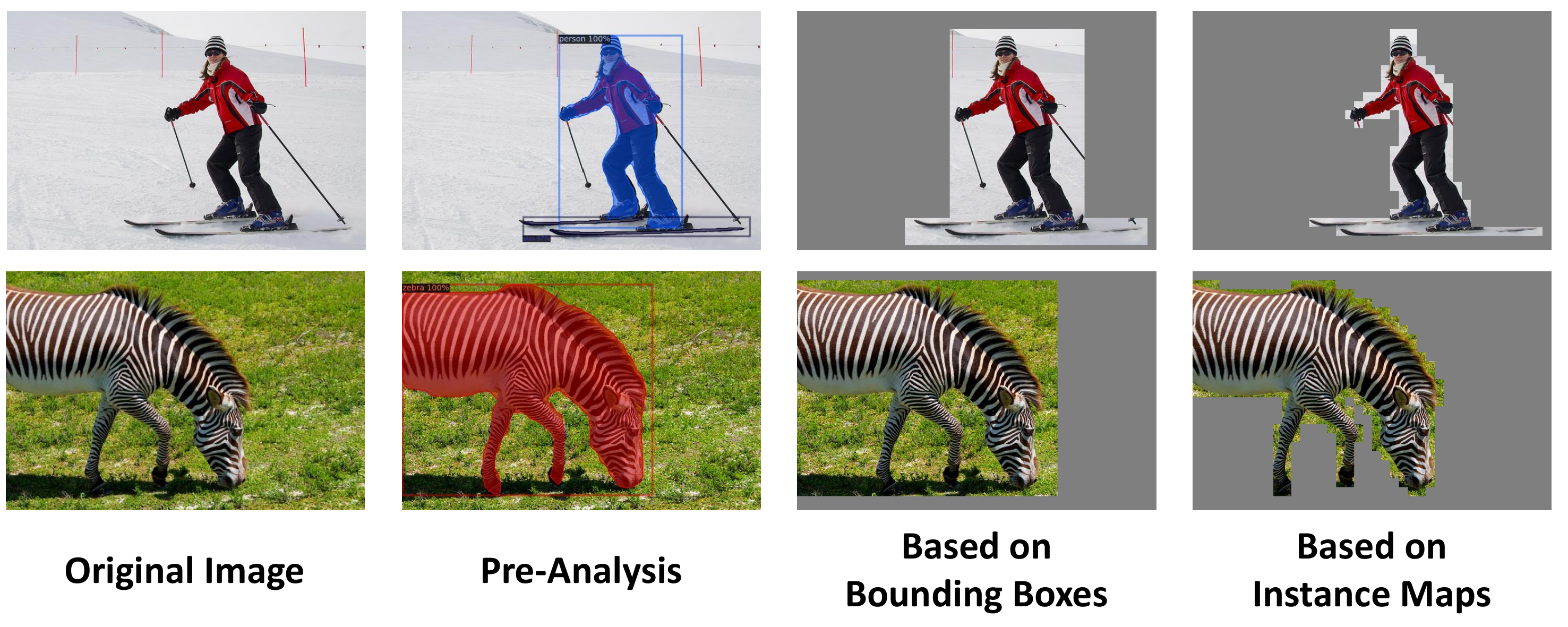}}
    \vspace{-3mm}

    \caption{Selective reconstructions from the semantically structured bitstream. The bitstream can be generated based on different pre-analysis methods. }
\label{fig:visualize_flexibility}
\vspace{-3mm}
\end{figure}

\myparagraph{Customizablity and Flexibility.} 
It's flexible to customize the semantically structured bitstream based on different semantical partition criteria. 
For instance, as shown in Figure \ref{fig:visualize_flexibility}, the generation and usage of semantically structured bitstreams can be based on either object detection or instance segmentation, depending on the specific requirements.
It should be noted that the methods of pre-analysis are not limited to object detection and instance segmentation as shown in Figure \ref{fig:visualize_flexibility}, but can include other techniques such as saliency detection\cite{hou2007saliency,goferman2011context,zhao2015saliency}, semantic segmentation\cite{xie2021segformer,zheng2021rethinking,zhao2017pyramid,long2015fully,badrinarayanan2017segnet}, panoptic segmentation\cite{cheng2020panoptic,kirillov2019panoptic}, \etc, and even human annotation.

\begin{figure}[t]
\centerline{\includegraphics[width=1.0\linewidth]{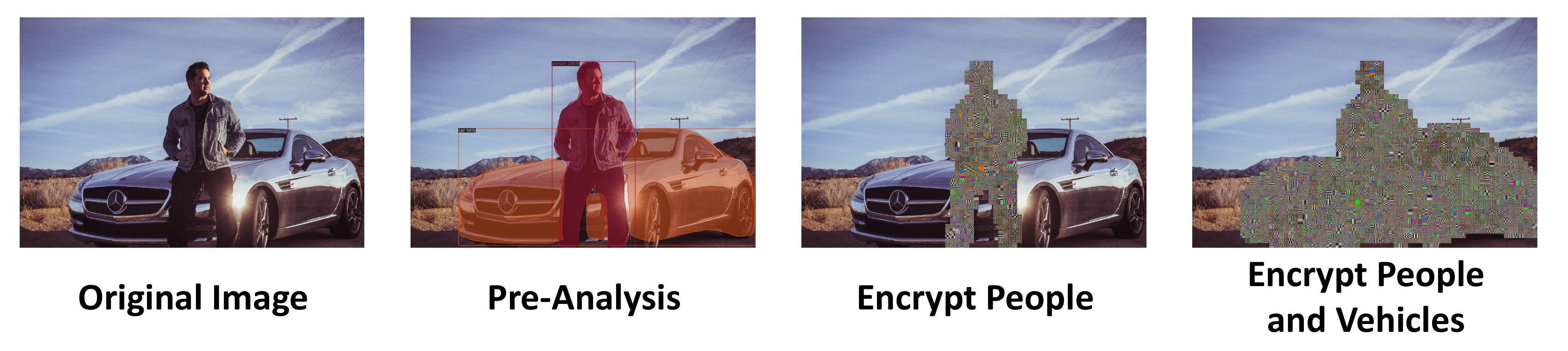}}
    \vspace{-3.5mm}

    \caption{Illustration of semantically-aware encryption based on our proposed method.}
\label{fig:visualize_encryption}
\vspace{-4mm}
\end{figure}

\myparagraph{Semantically-Aware Encryption.} 
As presented in Figure \ref{fig:visualize_encryption}, our proposed method can be applied to bitstream encryption, enabling selective and even layered encryption with semantic priors based on the user's security level. 
In some cases, the selectively encrypted SSB can allow for secure transmission and storage of sensitive information while minimizing the impact on visual quality and downstream analytics. 
Please refer to the supplementary material for details of implementation.

\section{Ablation Study}
In order to show the effectiveness of our proposed group mask and group-independent transform (GIT) for both RoI-aware selective reconstruction and downstream intelligent analytics, we conducted the ablation study at different bitrates. Specifically, we use people as the region of interest for RoI reconstruction, and the results for taking all foreground objects as RoI are presented in the supplementary material. 
Additionally, we conducted an ablation study on pose estimation of COCO 2017 for intelligent task support, and the results on instance segmentation are also included in the supplementary material.

\myparagraph{Group Mask.} 
Figure \ref{fig:Ablation} shows that group mask based partitioning saves significant bitrate compared with naive bounding box based partitioning. 
Meanwhile, both the visual quality of RoI reconstruction and downstream task performances would not be impeded.

\begin{figure}[t]
\vspace{-3mm}
  \centering
  \setlength{\abovecaptionskip}{2pt}
  \begin{subfigure}[b]{0.23\textwidth}
    \setlength{\abovecaptionskip}{-1pt}
    \includegraphics[width=\textwidth]{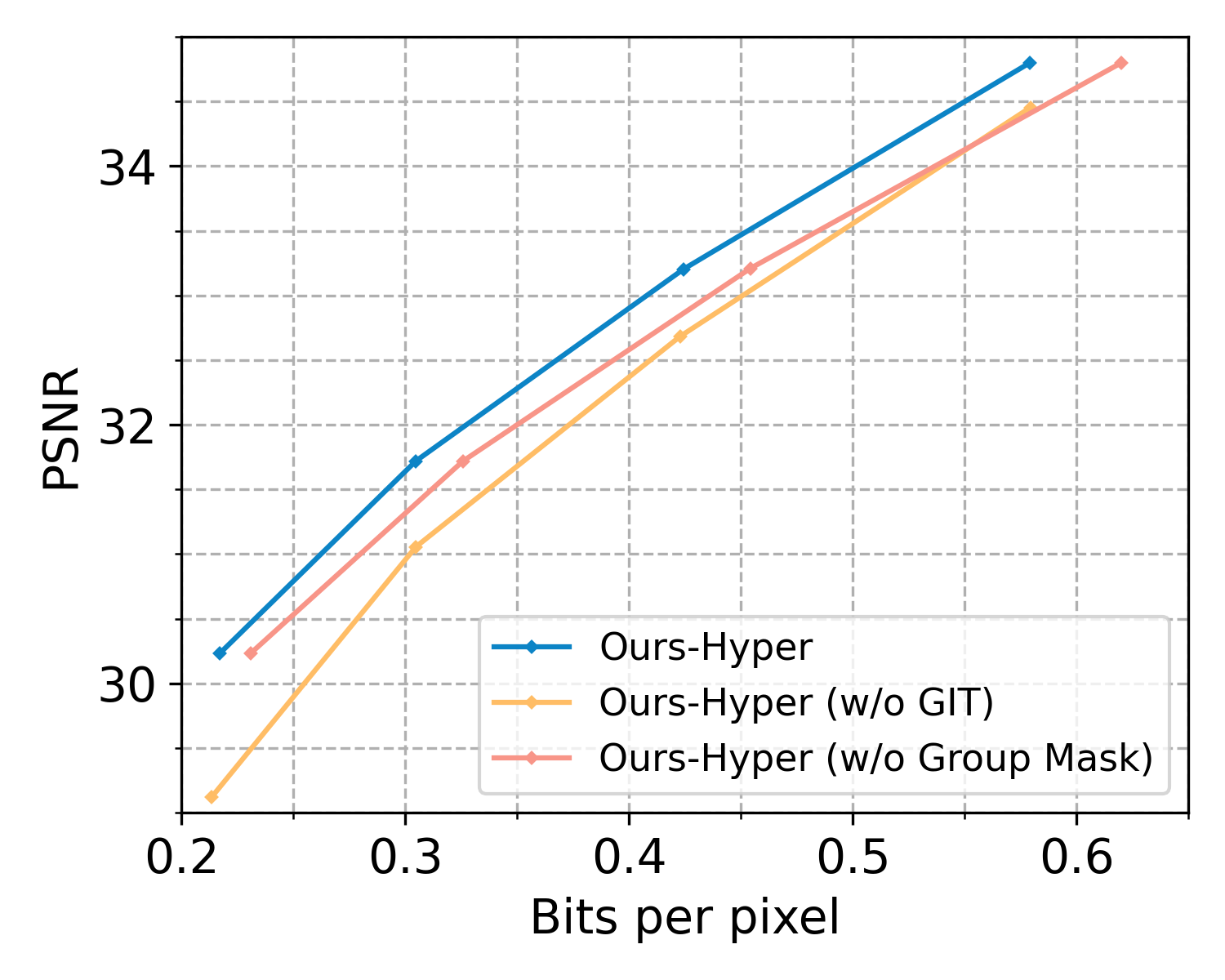}
    \caption{RoI reconstruction.}
    \label{fig:Ablation_humanrecon}
  \end{subfigure}
  \hspace{0mm}
  \begin{subfigure}[b]{0.23\textwidth}
    \setlength{\abovecaptionskip}{-1pt}    
    \includegraphics[width=\textwidth]{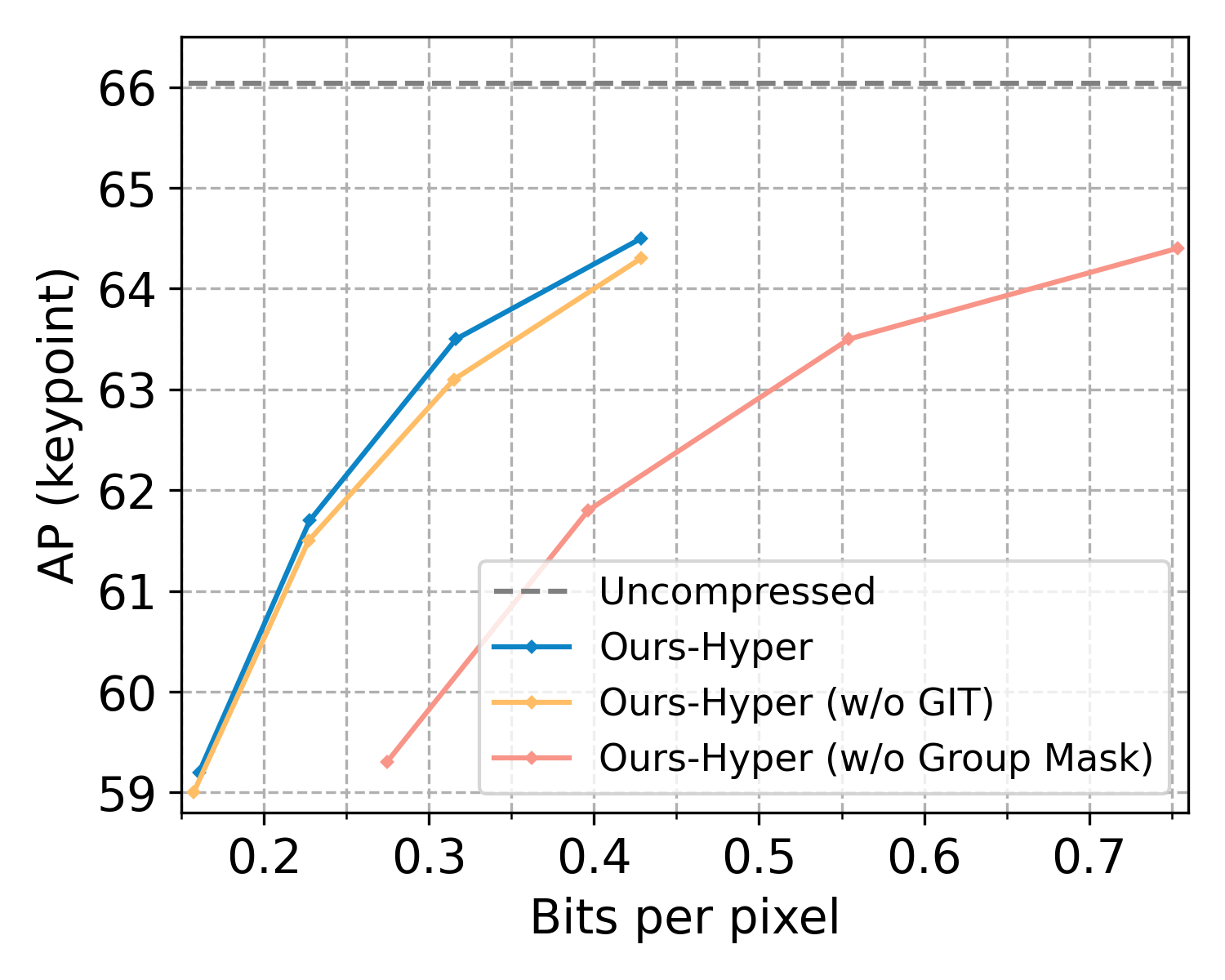}
    \caption{Intelligent task supporting.}
    \label{fig:Ablation_PoseEstimation}
  \end{subfigure}
  \caption{Ablation study.}
  \label{fig:Ablation}
  \vspace{-5mm}
\end{figure}

\myparagraph{Group-Independent transform.} 
As shown in Figure \ref{fig:Ablation_humanrecon}, group-independent transform is crucial for the visual quality of RoI reconstruction. 
And the downstream task performance would be impeded by the blurred and distorted region boundaries without group-independent transform, which is shown in Figure \ref{fig:Ablation_PoseEstimation}.

\section{Conclusion}
In this work, we propose to generate semantically structured bitstream with strong functionality based on group masks, which are highly flexible, customizable, while maintaining lightweight overhead. Moreover, we first propose the concept of group-independent transform and instantiate it by designing the Group-Independent Swin-Block (GI-Swin Block) to ensure independence among distinct groups, resulting in pleasing reconstructions of RoI regions. Particularly, our proposed method outperforms VTM-18.2 and achieves comparable coding efficiency with the SOTA neural image codecs, while offering strong functionality. 
The experimental results have demonstrated the effectiveness of our proposed method across several applications, including human consumption of regions of interest, downstream task support, and semantically-aware encryption. 


{\small
\bibliographystyle{ieee_fullname}
\bibliography{reference}

\begin{thebibliography}{10}\itemsep=-1pt

\bibitem{agustsson2017soft}
Eirikur Agustsson, Fabian Mentzer, Michael Tschannen, Lukas Cavigelli, Radu
  Timofte, Luca Benini, and Luc~V Gool.
\newblock Soft-to-hard vector quantization for end-to-end learning compressible
  representations.
\newblock {\em NeurIPS}, 30, 2017.

\bibitem{agustsson2019generative}
Eirikur Agustsson, Michael Tschannen, Fabian Mentzer, Radu Timofte, and Luc~Van
  Gool.
\newblock Generative adversarial networks for extreme learned image
  compression.
\newblock In {\em ICCV}, pages 221--231, 2019.

\bibitem{akbari2019dsslic}
Mohammad Akbari, Jie Liang, and Jingning Han.
\newblock Dsslic: Deep semantic segmentation-based layered image compression.
\newblock In {\em ICASSP 2019-2019 IEEE International Conference on Acoustics,
  Speech and Signal Processing (ICASSP)}, pages 2042--2046. IEEE, 2019.

\bibitem{babenko2014neural}
Artem Babenko, Anton Slesarev, Alexandr Chigorin, and Victor Lempitsky.
\newblock Neural codes for image retrieval.
\newblock In {\em ECCV}, pages 584--599. Springer, 2014.

\bibitem{badrinarayanan2017segnet}
Vijay Badrinarayanan, Alex Kendall, and Roberto Cipolla.
\newblock Segnet: A deep convolutional encoder-decoder architecture for image
  segmentation.
\newblock {\em TPAMI}, 39(12):2481--2495, 2017.

\bibitem{balle2020nonlinear}
Johannes Ball{\'e}, Philip~A Chou, David Minnen, Saurabh Singh, Nick Johnston,
  Eirikur Agustsson, Sung~Jin Hwang, and George Toderici.
\newblock Nonlinear transform coding.
\newblock {\em IEEE Journal of Selected Topics in Signal Processing},
  15(2):339--353, 2020.

\bibitem{balle2017end}
Johannes Ball{\'e}, Valero Laparra, and Eero~P Simoncelli.
\newblock End-to-end optimized image compression.
\newblock In {\em ICLR}, 2017.

\bibitem{balle2018variational}
Johannes Ball{\'e}, David Minnen, Saurabh Singh, Sung~Jin Hwang, and Nick
  Johnston.
\newblock Variational image compression with a scale hyperprior.
\newblock In {\em ICLR}, 2018.

\bibitem{bolya2019yolact}
Daniel Bolya, Chong Zhou, Fanyi Xiao, and Yong~Jae Lee.
\newblock Yolact: Real-time instance segmentation.
\newblock In {\em ICCV}, pages 9157--9166, 2019.

\bibitem{bross2021overview}
Benjamin Bross, Ye-Kui Wang, Yan Ye, Shan Liu, Jianle Chen, Gary~J Sullivan,
  and Jens-Rainer Ohm.
\newblock Overview of the versatile video coding (vvc) standard and its
  applications.
\newblock {\em TCSVT}, 2021.

\bibitem{chen2017deeplab}
Liang-Chieh Chen, George Papandreou, Iasonas Kokkinos, Kevin Murphy, and Alan~L
  Yuille.
\newblock Deeplab: Semantic image segmentation with deep convolutional nets,
  atrous convolution, and fully connected crfs.
\newblock {\em TPAMI}, 40(4):834--848, 2017.

\bibitem{chen2018encoder}
Liang-Chieh Chen, Yukun Zhu, George Papandreou, Florian Schroff, and Hartwig
  Adam.
\newblock Encoder-decoder with atrous separable convolution for semantic image
  segmentation.
\newblock In {\em ECCV}, pages 801--818, 2018.

\bibitem{chen2019toward}
Zhuo Chen, Kui Fan, Shiqi Wang, Lingyu Duan, Weisi Lin, and Alex~Chichung Kot.
\newblock Toward intelligent sensing: Intermediate deep feature compression.
\newblock {\em TIP}, 29:2230--2243, 2019.

\bibitem{chen2019lossy}
Zhuo Chen, Kui Fan, Shiqi Wang, Ling-Yu Duan, Weisi Lin, and Alex Kot.
\newblock Lossy intermediate deep learning feature compression and evaluation.
\newblock In {\em ACM MM}, pages 2414--2422, 2019.

\bibitem{cheng2020panoptic}
Bowen Cheng, Maxwell~D Collins, Yukun Zhu, Ting Liu, Thomas~S Huang, Hartwig
  Adam, and Liang-Chieh Chen.
\newblock Panoptic-deeplab: A simple, strong, and fast baseline for bottom-up
  panoptic segmentation.
\newblock In {\em CVPR}, pages 12475--12485, 2020.

\bibitem{cheng2020learned}
Zhengxue Cheng, Heming Sun, Masaru Takeuchi, and Jiro Katto.
\newblock Learned image compression with discretized gaussian mixture
  likelihoods and attention modules.
\newblock In {\em CVPR}, pages 7939--7948, 2020.

\bibitem{dosovitskiy2020image}
Alexey Dosovitskiy, Lucas Beyer, Alexander Kolesnikov, Dirk Weissenborn,
  Xiaohua Zhai, Thomas Unterthiner, Mostafa Dehghani, Matthias Minderer, Georg
  Heigold, Sylvain Gelly, et~al.
\newblock An image is worth 16x16 words: Transformers for image recognition at
  scale.
\newblock {\em ICLR}, 2020.

\bibitem{duan2015overview}
Ling-Yu Duan, Vijay Chandrasekhar, Jie Chen, Jie Lin, Zhe Wang, Tiejun Huang,
  Bernd Girod, and Wen Gao.
\newblock Overview of the mpeg-cdvs standard.
\newblock {\em TIP}, 25(1):179--194, 2015.

\bibitem{duan2018compact}
Ling-Yu Duan, Yihang Lou, Yan Bai, Tiejun Huang, Wen Gao, Vijay Chandrasekhar,
  Jie Lin, Shiqi Wang, and Alex~Chichung Kot.
\newblock Compact descriptors for video analysis: The emerging mpeg standard.
\newblock {\em IEEE MultiMedia}, 26(2):44--54, 2018.

\bibitem{ebrahimi2000mpeg}
Touradj Ebrahimi and Caspar Horne.
\newblock Mpeg-4 natural video coding--an overview.
\newblock {\em Signal Processing: Image Communication}, 15(4-5):365--385, 2000.

\bibitem{feng2022image}
Ruoyu Feng, Xin Jin, Zongyu Guo, Runsen Feng, Yixin Gao, Tianyu He, Zhizheng
  Zhang, Simeng Sun, and Zhibo Chen.
\newblock Image coding for machines with omnipotent feature learning.
\newblock In {\em ECCV}, pages 510--528. Springer, 2022.

\bibitem{fisher1963statistical}
Ronald~Aylmer Fisher, Frank Yates, et~al.
\newblock {\em Statistical tables for biological, agricultural and medical
  research, edited by ra fisher and f. yates}.
\newblock Edinburgh: Oliver and Boyd, 1963.

\bibitem{goferman2011context}
Stas Goferman, Lihi Zelnik-Manor, and Ayellet Tal.
\newblock Context-aware saliency detection.
\newblock {\em TPAMI}, 34(10):1915--1926, 2011.

\bibitem{goyal2001theoretical}
Vivek~K Goyal.
\newblock Theoretical foundations of transform coding.
\newblock {\em IEEE Signal Processing Magazine}, 18(5):9--21, 2001.

\bibitem{guo2021causal}
Zongyu Guo, Zhizheng Zhang, Runsen Feng, and Zhibo Chen.
\newblock Causal contextual prediction for learned image compression.
\newblock {\em TCSVT}, 32(4):2329--2341, 2021.

\bibitem{guo2021soft}
Zongyu Guo, Zhizheng Zhang, Runsen Feng, and Zhibo Chen.
\newblock Soft then hard: Rethinking the quantization in neural image
  compression.
\newblock In {\em ICML}, pages 3920--3929. PMLR, 2021.

\bibitem{han2021transformer}
Kai Han, An Xiao, Enhua Wu, Jianyuan Guo, Chunjing Xu, and Yunhe Wang.
\newblock Transformer in transformer.
\newblock {\em NeurIPS}, 34:15908--15919, 2021.

\bibitem{he2022elic}
Dailan He, Ziming Yang, Weikun Peng, Rui Ma, Hongwei Qin, and Yan Wang.
\newblock Elic: Efficient learned image compression with unevenly grouped
  space-channel contextual adaptive coding.
\newblock In {\em CVPR}, pages 5718--5727, 2022.

\bibitem{he2017mask}
Kaiming He, Georgia Gkioxari, Piotr Doll{\'a}r, and Ross Girshick.
\newblock Mask r-cnn.
\newblock In {\em ICCV}, pages 2961--2969, 2017.

\bibitem{he2016deep}
Kaiming He, Xiangyu Zhang, Shaoqing Ren, and Jian Sun.
\newblock Deep residual learning for image recognition.
\newblock In {\em CVPR}, pages 770--778, 2016.

\bibitem{hou2007saliency}
Xiaodi Hou and Liqing Zhang.
\newblock Saliency detection: A spectral residual approach.
\newblock In {\em CVPR}, pages 1--8. Ieee, 2007.

\bibitem{hu2020towards}
Yueyu Hu, Shuai Yang, Wenhan Yang, Ling-Yu Duan, and Jiaying Liu.
\newblock Towards coding for human and machine vision: A scalable image coding
  approach.
\newblock In {\em 2020 IEEE International Conference on Multimedia and Expo
  (ICME)}, pages 1--6. IEEE, 2020.

\bibitem{huang2021visual}
Zhimeng Huang, Chuanmin Jia, Shanshe Wang, and Siwei Ma.
\newblock Visual analysis motivated rate-distortion model for image coding.
\newblock In {\em 2021 IEEE International Conference on Multimedia and Expo
  (ICME)}, pages 1--6. IEEE, 2021.

\bibitem{jia2022visual}
Menglin Jia, Luming Tang, Bor-Chun Chen, Claire Cardie, Serge Belongie, Bharath
  Hariharan, and Ser-Nam Lim.
\newblock Visual prompt tuning.
\newblock {\em ECCV}, 2022.

\bibitem{jin2022semantically}
Xin Jin, Ruoyu Feng, Simeng Sun, Runsen Feng, Tianyu He, and Zhibo Chen.
\newblock Semantically video coding: Instill static-dynamic clues into
  structured bitstream for ai tasks.
\newblock {\em arXiv preprint arXiv:2201.10162}, 2022.

\bibitem{johnston2018improved}
Nick Johnston, Damien Vincent, David Minnen, Michele Covell, Saurabh Singh,
  Troy Chinen, Sung~Jin Hwang, Joel Shor, and George Toderici.
\newblock Improved lossy image compression with priming and spatially adaptive
  bit rates for recurrent networks.
\newblock In {\em CVPR}, pages 4385--4393, 2018.

\bibitem{katsaggelos1998mpeg}
Aggelos~K Katsaggelos, Lisimachos~P Kondi, Fabian~W Meier, J{\"o}rn Ostermann,
  and Guido~M Schuster.
\newblock Mpeg-4 and rate-distortion-based shape-coding techniques.
\newblock {\em Proceedings of the IEEE}, 86(6):1126--1154, 1998.

\bibitem{kirillov2019panoptic}
Alexander Kirillov, Kaiming He, Ross Girshick, Carsten Rother, and Piotr
  Doll{\'a}r.
\newblock Panoptic segmentation.
\newblock In {\em CVPR}, pages 9404--9413, 2019.

\bibitem{kodak}
Eastman Kodak.
\newblock Kodak lossless true color image suite (photocd pcd0992).
\newblock \url{http://r0k.us/graphics/kodak}, 1993.

\bibitem{le2021image}
Nam Le, Honglei Zhang, Francesco Cricri, Ramin Ghaznavi-Youvalari, and Esa
  Rahtu.
\newblock Image coding for machines: An end-to-end learned approach.
\newblock In {\em ICASSP 2021-2021 IEEE International Conference on Acoustics,
  Speech and Signal Processing (ICASSP)}, pages 1590--1594. IEEE, 2021.

\bibitem{le2021learned}
Nam Le, Honglei Zhang, Francesco Cricri, Ramin Ghaznavi-Youvalari,
  Hamed~Rezazadegan Tavakoli, and Esa Rahtu.
\newblock Learned image coding for machines: A content-adaptive approach.
\newblock In {\em 2021 IEEE International Conference on Multimedia and Expo
  (ICME)}, pages 1--6. IEEE, 2021.

\bibitem{li2020learning}
Mu Li, Wangmeng Zuo, Shuhang Gu, Jane You, and David Zhang.
\newblock Learning content-weighted deep image compression.
\newblock {\em TPAMI}, 2020.

\bibitem{li2018learning}
Mu Li, Wangmeng Zuo, Shuhang Gu, Debin Zhao, and David Zhang.
\newblock Learning convolutional networks for content-weighted image
  compression.
\newblock In {\em CVPR}, pages 3214--3223, 2018.

\bibitem{li2021task}
Xin Li, Jun Shi, and Zhibo Chen.
\newblock Task-driven semantic coding via reinforcement learning.
\newblock {\em TIP}, 2021.

\bibitem{li2022exploring}
Yanghao Li, Hanzi Mao, Ross Girshick, and Kaiming He.
\newblock Exploring plain vision transformer backbones for object detection.
\newblock {\em ECCV}, 2022.

\bibitem{lin2017feature}
Tsung-Yi Lin, Piotr Doll{\'a}r, Ross Girshick, Kaiming He, Bharath Hariharan,
  and Serge Belongie.
\newblock Feature pyramid networks for object detection.
\newblock In {\em CVPR}, pages 2117--2125, 2017.

\bibitem{lin2014microsoft}
Tsung-Yi Lin, Michael Maire, Serge Belongie, James Hays, Pietro Perona, Deva
  Ramanan, Piotr Doll{\'a}r, and C~Lawrence Zitnick.
\newblock Microsoft coco: Common objects in context.
\newblock In {\em ECCV}, pages 740--755. Springer, 2014.

\bibitem{liu2018path}
Shu Liu, Lu Qi, Haifang Qin, Jianping Shi, and Jiaya Jia.
\newblock Path aggregation network for instance segmentation.
\newblock In {\em CVPR}, pages 8759--8768, 2018.

\bibitem{liu2021swin}
Ze Liu, Yutong Lin, Yue Cao, Han Hu, Yixuan Wei, Zheng Zhang, Stephen Lin, and
  Baining Guo.
\newblock Swin transformer: Hierarchical vision transformer using shifted
  windows.
\newblock In {\em ICCV}, pages 10012--10022, 2021.

\bibitem{long2015fully}
Jonathan Long, Evan Shelhamer, and Trevor Darrell.
\newblock Fully convolutional networks for semantic segmentation.
\newblock In {\em CVPR}, pages 3431--3440, 2015.

\bibitem{ma2018joint}
Siwei Ma, Xiang Zhang, Shiqi Wang, Xinfeng Zhang, Chuanmin Jia, and Shanshe
  Wang.
\newblock Joint feature and texture coding: Toward smart video representation
  via front-end intelligence.
\newblock {\em TCSVT}, 29(10):3095--3105, 2018.

\bibitem{mentzer2018conditional}
Fabian Mentzer, Eirikur Agustsson, Michael Tschannen, Radu Timofte, and Luc
  Van~Gool.
\newblock Conditional probability models for deep image compression.
\newblock In {\em CVPR}, pages 4394--4402, 2018.

\bibitem{mentzer2020high}
Fabian Mentzer, George Toderici, Michael Tschannen, and Eirikur Agustsson.
\newblock High-fidelity generative image compression.
\newblock {\em arXiv preprint arXiv:2006.09965}, 2020.

\bibitem{minnen2018joint}
David Minnen, Johannes Ball{\'e}, and George Toderici.
\newblock Joint autoregressive and hierarchical priors for learned image
  compression.
\newblock In {\em NeurIPS}, 2018.

\bibitem{minnen2020channel}
David Minnen and Saurabh Singh.
\newblock Channel-wise autoregressive entropy models for learned image
  compression.
\newblock In {\em 2020 IEEE International Conference on Image Processing
  (ICIP)}, pages 3339--3343. IEEE, 2020.

\bibitem{rabbani2002overview}
Majid Rabbani and Rajan Joshi.
\newblock An overview of the jpeg 2000 still image compression standard.
\newblock {\em Signal processing: Image communication}, 17(1):3--48, 2002.

\bibitem{redmon2016you}
Joseph Redmon, Santosh Divvala, Ross Girshick, and Ali Farhadi.
\newblock You only look once: Unified, real-time object detection.
\newblock In {\em CVPR}, pages 779--788, 2016.

\bibitem{redmon2017yolo9000}
Joseph Redmon and Ali Farhadi.
\newblock Yolo9000: better, faster, stronger.
\newblock In {\em CVPR}, pages 7263--7271, 2017.

\bibitem{ren2015faster}
Shaoqing Ren, Kaiming He, Ross Girshick, and Jian Sun.
\newblock Faster r-cnn: Towards real-time object detection with region proposal
  networks.
\newblock {\em NeurIPS}, 28:91--99, 2015.

\bibitem{rippel2017real}
Oren Rippel and Lubomir Bourdev.
\newblock Real-time adaptive image compression.
\newblock In {\em ICML}, pages 2922--2930. PMLR, 2017.

\bibitem{sikora1997mpeg}
Thomas Sikora.
\newblock The mpeg-4 video standard verification model.
\newblock {\em TCSVT}, 7(1):19--31, 1997.

\bibitem{singh2020end}
Saurabh Singh, Sami Abu-El-Haija, Nick Johnston, Johannes Ball{\'e}, Abhinav
  Shrivastava, and George Toderici.
\newblock End-to-end learning of compressible features.
\newblock In {\em 2020 IEEE International Conference on Image Processing
  (ICIP)}, pages 3349--3353. IEEE, 2020.

\bibitem{sullivan2012overview}
Gary~J Sullivan, Jens-Rainer Ohm, Woo-Jin Han, and Thomas Wiegand.
\newblock Overview of the high efficiency video coding (hevc) standard.
\newblock {\em TCSVT}, 22(12):1649--1668, 2012.

\bibitem{sun2020semantic}
Simeng Sun, Tianyu He, and Zhibo Chen.
\newblock Semantic structured image coding framework for multiple intelligent
  applications.
\newblock {\em TCSVT}, 2020.

\bibitem{theis2017lossy}
Lucas Theis, Wenzhe Shi, Andrew Cunningham, and Ferenc Husz{\'a}r.
\newblock Lossy image compression with compressive autoencoders.
\newblock {\em arXiv preprint arXiv:1703.00395}, 2017.

\bibitem{vaswani2017attention}
Ashish Vaswani, Noam Shazeer, Niki Parmar, Jakob Uszkoreit, Llion Jones,
  Aidan~N Gomez, {\L}ukasz Kaiser, and Illia Polosukhin.
\newblock Attention is all you need.
\newblock {\em NeurIPS}, 30, 2017.

\bibitem{wallace1992jpeg}
Gregory~K Wallace.
\newblock The jpeg still picture compression standard.
\newblock {\em IEEE transactions on consumer electronics}, 38(1):xviii--xxxiv,
  1992.

\bibitem{wang2021towards}
Shurun Wang, Shiqi Wang, Wenhan Yang, Xinfeng Zhang, Shanshe Wang, Siwei Ma,
  and Wen Gao.
\newblock Towards analysis-friendly face representation with scalable feature
  and texture compression.
\newblock {\em TMM}, 2021.

\bibitem{wiegand2003overview}
Thomas Wiegand, Gary~J Sullivan, Gisle Bjontegaard, and Ajay Luthra.
\newblock Overview of the h. 264/avc video coding standard.
\newblock {\em TCSVT}, 13(7):560--576, 2003.

\bibitem{wu2019detectron2}
Yuxin Wu, Alexander Kirillov, Francisco Massa, Wan-Yen Lo, and Ross Girshick.
\newblock Detectron2.
\newblock \url{https://github.com/facebookresearch/detectron2}, 2019.

\bibitem{xie2021segformer}
Enze Xie, Wenhai Wang, Zhiding Yu, Anima Anandkumar, Jose~M Alvarez, and Ping
  Luo.
\newblock Segformer: Simple and efficient design for semantic segmentation with
  transformers.
\newblock {\em NeurIPS}, 34:12077--12090, 2021.

\bibitem{yang2020improving}
Yibo Yang, Robert Bamler, and Stephan Mandt.
\newblock Improving inference for neural image compression.
\newblock {\em NeurIPS}, 33:573--584, 2020.

\bibitem{zhao2017pyramid}
Hengshuang Zhao, Jianping Shi, Xiaojuan Qi, Xiaogang Wang, and Jiaya Jia.
\newblock Pyramid scene parsing network.
\newblock In {\em CVPR}, pages 2881--2890, 2017.

\bibitem{zhao2015saliency}
Rui Zhao, Wanli Ouyang, Hongsheng Li, and Xiaogang Wang.
\newblock Saliency detection by multi-context deep learning.
\newblock In {\em CVPR}, pages 1265--1274, 2015.

\bibitem{zheng2021rethinking}
Sixiao Zheng, Jiachen Lu, Hengshuang Zhao, Xiatian Zhu, Zekun Luo, Yabiao Wang,
  Yanwei Fu, Jianfeng Feng, Tao Xiang, Philip~HS Torr, et~al.
\newblock Rethinking semantic segmentation from a sequence-to-sequence
  perspective with transformers.
\newblock In {\em CVPR}, pages 6881--6890, 2021.

\bibitem{zhou2019end}
Lei Zhou, Zhenhong Sun, Xiangji Wu, and Junmin Wu.
\newblock End-to-end optimized image compression with attention mechanism.
\newblock In {\em CVPR workshops}, page~0, 2019.

\bibitem{zhu2022transformer}
Yinhao Zhu, Yang Yang, and Taco Cohen.
\newblock Transformer-based transform coding.
\newblock In {\em ICLR}, 2022.

\end{thebibliography}
}

\clearpage
\clearpage

\newpage
\appendix
This supplementary material provides additional experimental comparison, implementation details of semantically-aware encryption, and ablation study of taking all foreground objects as regions of interest.

\section{Comparison with Other Methods}
In this section, we compare our proposed method with VTM 18.2 and three other methods designed for image coding for machines: the traditional codec based RoI bit-allocation scheme\cite{huang2021visual}, the learning based joint training codec\cite{le2021image}, and the general representation learning based approach \cite{feng2022image}. 
The evaluation task is instance segmentation on COCO 2017\cite{lin2014microsoft}.
It is noteworthy that the bitstreams of RoI bit-allocation\cite{huang2021visual} and the task-driven joint training methods are tailored to the corresponding task, and the general representation learning based method\cite{feng2022image} necessitates retraining the task model with new source data (i.e., the learned representation).
Nevertheless, our method still yields significant improvement compared with other methods.

\begin{figure}[h!]
\setlength{\abovecaptionskip}{0pt}
\setlength{\belowcaptionskip}{0pt}
\centerline{\includegraphics[width=0.7\linewidth]{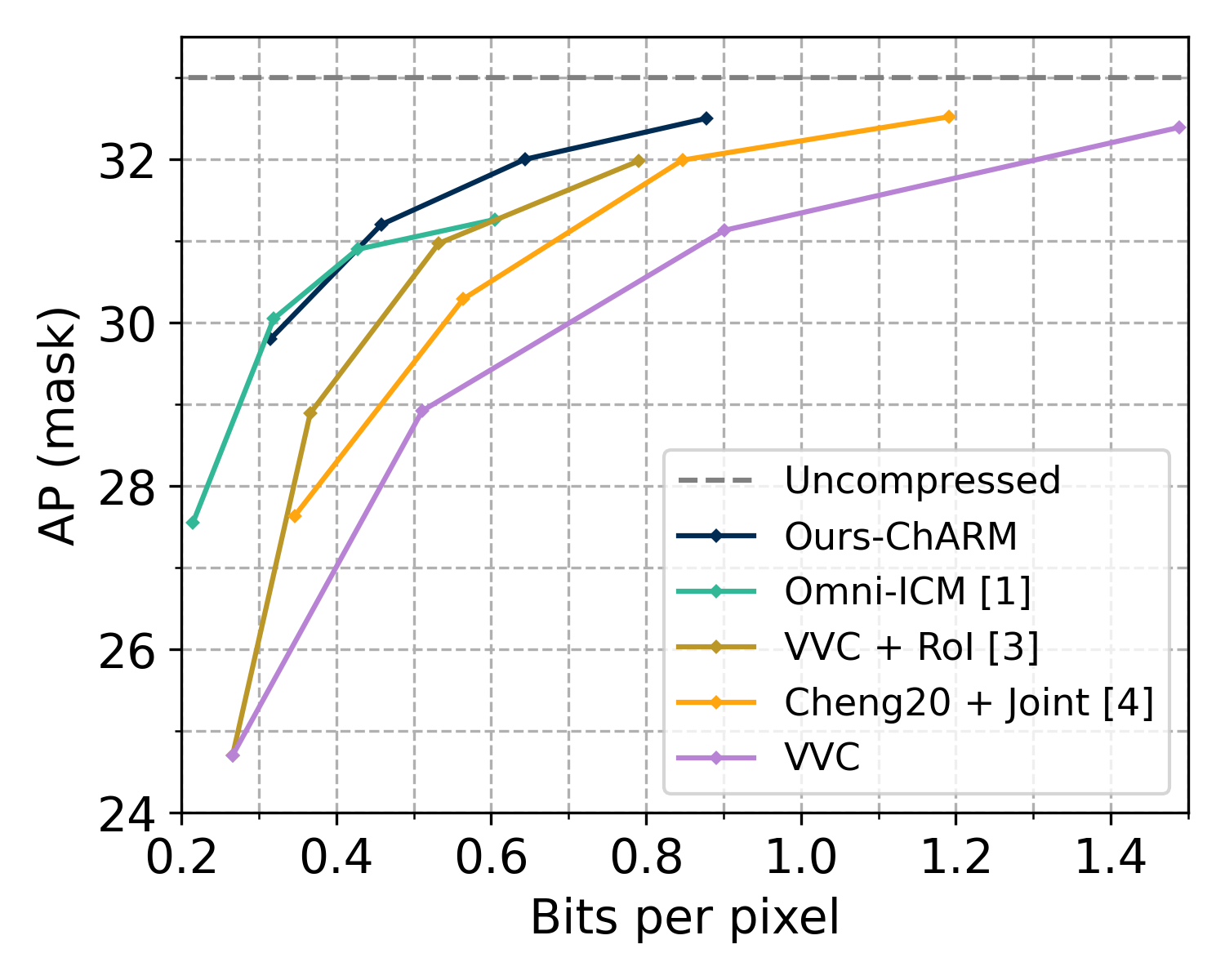}}
    \caption{
    Instance segmentation on MS COCO.
    }
\label{fig:ComparedwithotherICMmethods_InstanceSegmentation}
\end{figure}

\section{Semantically-Aware Encryption}
We conduct encryption on latent variables. 
Specifically, for each group to be encrypted, the chosen variables are first reshaped into a one dimensional vector, and then perturbed by Fisher-Yates shuffle algorithm\cite{fisher1963statistical}. 
The perturbance is achieved through a cryptographic random seed.
And, the perturbed vector is rearranged into its original shape and encoded into the corresponding bitstream. 
Correspondingly, the likelihood is shuffled in the same way for entropy coding. 
Then, entropy coding are conducted on the latent variables of current group, resulting in the encrypted bitstream. 
At last, the semantically structured bitstream consists of both encrypted and non-encrypted ones according to requirements.
Furthermore, the user can perform different levels of encryption on different groups, resulting in the layered encrypted bitstream.

\section{Ablation Study}
\begin{figure}[h]
\vspace{-3mm}
  \centering
  \setlength{\abovecaptionskip}{2pt}
  \begin{subfigure}[b]{0.23\textwidth}
    \setlength{\abovecaptionskip}{-1pt}
    \includegraphics[width=\textwidth]{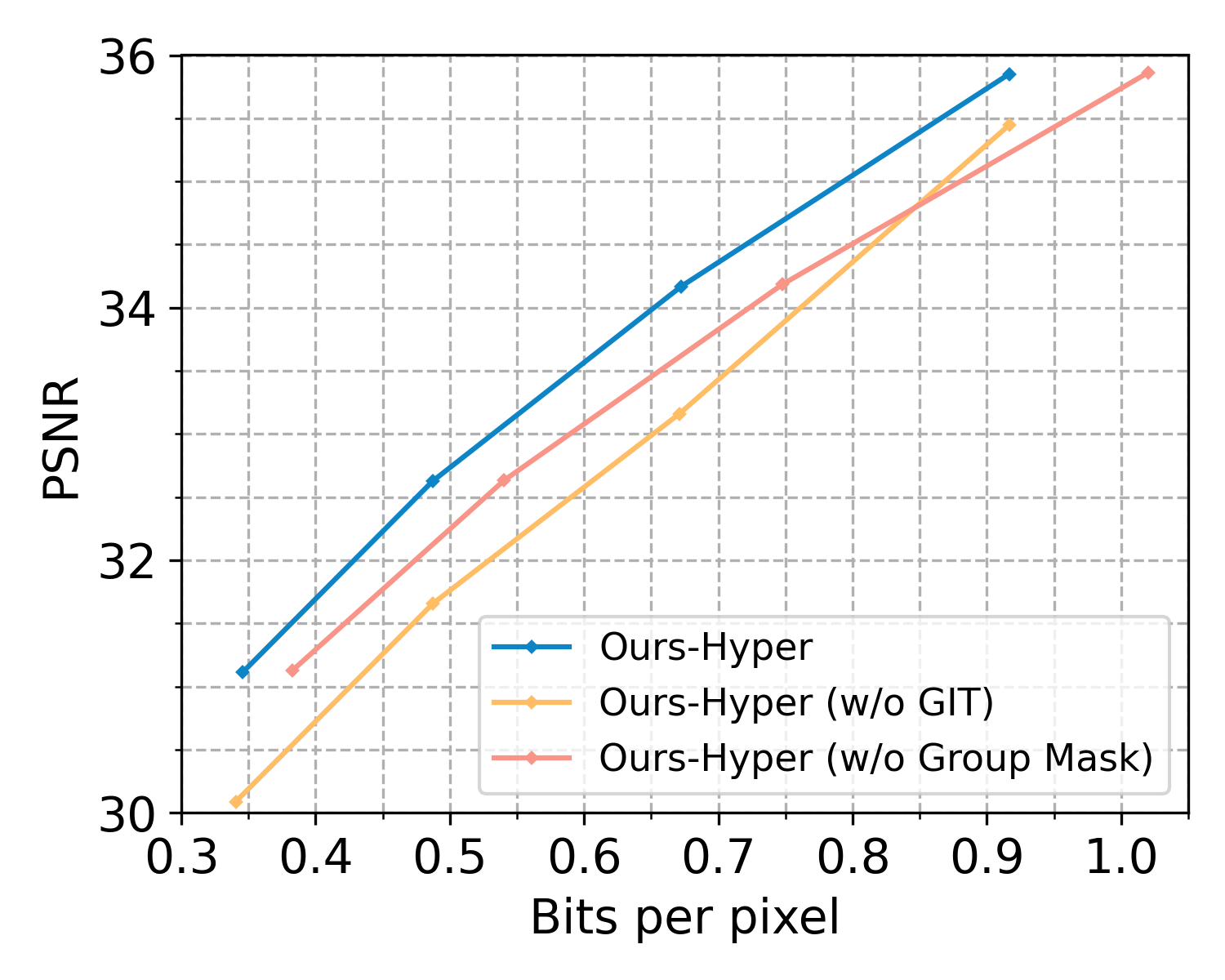}
    \caption{RoI reconstruction.}
    \label{fig:Ablation_humanrecon}
  \end{subfigure}
  \hspace{0mm}
  \begin{subfigure}[b]{0.23\textwidth}
    \setlength{\abovecaptionskip}{-1pt}    
    \includegraphics[width=\textwidth]{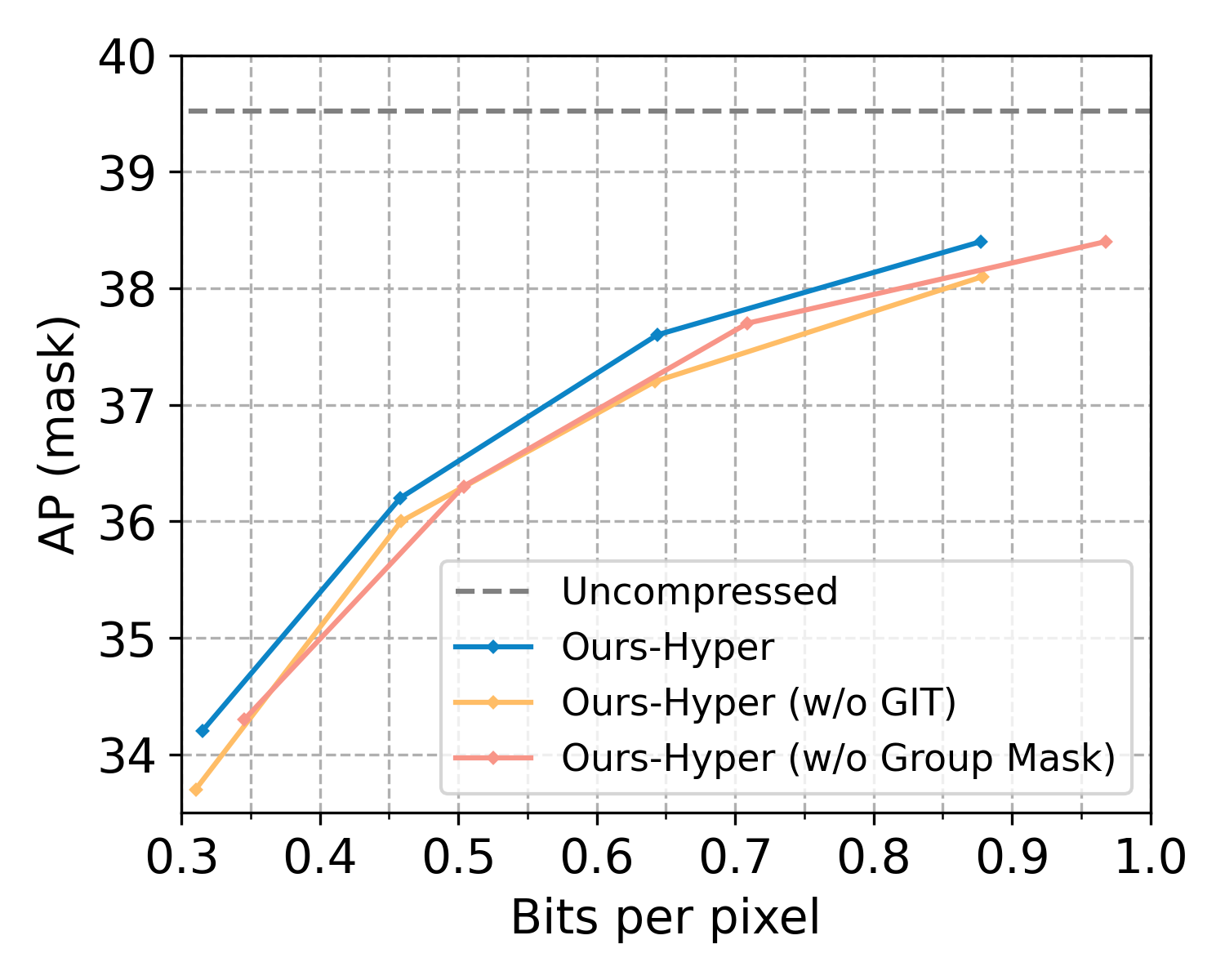}
    \caption{Instance Segmentation.}
    \label{fig:Ablation_PoseEstimation}
  \end{subfigure}
  \caption{Ablation study.}
  \label{fig:Ablation}
\end{figure}
In Section 5 of the main text, we perform ablation studies by regarding people as the regions of interest (RoI). In this section, we provide additional experimental results, where we consider all foreground objects belonging to MS COCO's categories as the RoI. 
The results shown in Figure~\ref{fig:Ablation} indicate the same conclusion as Section 5 of the main text, that our proposed group mask can significantly save bitrate while the group-independent transform can ensure the reconstruction quality of selective transmission and reconstruction.

\end{document}